\documentclass[aps,prx,twocolumn,longbibliography,superscriptaddress,floatfix,nofootinbib]{revtex4-2}

\usepackage{epsfig,amsmath,amssymb,color,physics}
\usepackage[makeroom]{cancel}
\usepackage[caption=false]{subfig}
\usepackage{mathrsfs}
\usepackage[countmax]{subfloat}
\usepackage[normalem]{ulem}
\usepackage[english]{babel}
\usepackage{dsfont}
\usepackage{float}
\usepackage[bookmarks=true,colorlinks,linkcolor=black,urlcolor=NavyBlue,citecolor=RoyalBlue]{hyperref}
\usepackage[dvipsnames]{xcolor}
\usepackage{bbold}
\usepackage{mathtools}
\usepackage{enumitem}
\usepackage{color}

\newcommand{\beginsupplement}{
    \setcounter{table}{0}
    \renewcommand{\thetable}{S\arabic{table}}
    \setcounter{figure}{0}
    \renewcommand{\thefigure}{S\arabic{figure}}
    \setcounter{equation}{0}
    \renewcommand{\theequation}{S\arabic{equation}}
}

\newcommand{\omg}{\overline{\omega}}

\begin{document}

\title{
Disorder-tunable entanglement at infinite temperature 
}

\author{Hang Dong}\thanks{These three authors contributed equally}
\affiliation{School of Physics, and ZJU-Hangzhou Global Scientific and Technological Innovation Center, Zhejiang University, Hangzhou 310027, China}
\author{Jean-Yves Desaules}\thanks{These three authors contributed equally}
\affiliation{School of Physics and Astronomy, University of Leeds, Leeds, UK}
\author{Yu Gao}\thanks{These three authors contributed equally}
\affiliation{School of Physics, and ZJU-Hangzhou Global Scientific and Technological Innovation Center, Zhejiang University, Hangzhou 310027, China}
\author{Ning Wang}
\affiliation{School of Physics, and ZJU-Hangzhou Global Scientific and Technological Innovation Center, Zhejiang University, Hangzhou 310027, China}
\author{Zexian Guo}
\affiliation{School of Physics, and ZJU-Hangzhou Global Scientific and Technological Innovation Center, Zhejiang University, Hangzhou 310027, China}
\author{Jiachen Chen}
\affiliation{School of Physics, and ZJU-Hangzhou Global Scientific and Technological Innovation Center, Zhejiang University, Hangzhou 310027, China}
\author{Yiren Zou}
\affiliation{School of Physics, and ZJU-Hangzhou Global Scientific and Technological Innovation Center, Zhejiang University, Hangzhou 310027, China}
\author{Feitong Jin}
\affiliation{School of Physics, and ZJU-Hangzhou Global Scientific and Technological Innovation Center, Zhejiang University, Hangzhou 310027, China}
\author{Xuhao Zhu}
\affiliation{School of Physics, and ZJU-Hangzhou Global Scientific and Technological Innovation Center, Zhejiang University, Hangzhou 310027, China}
\author{Pengfei Zhang}
\affiliation{School of Physics, and ZJU-Hangzhou Global Scientific and Technological Innovation Center, Zhejiang University, Hangzhou 310027, China}

\author{Hekang Li}
\affiliation{School of Physics, and ZJU-Hangzhou Global Scientific and Technological Innovation Center, Zhejiang University, Hangzhou 310027, China}
\author{Zhen Wang}\email{2010wangzhen@zju.edu.cn}
\affiliation{School of Physics, and ZJU-Hangzhou Global Scientific and Technological Innovation Center, Zhejiang University, Hangzhou 310027, China}
\author{Qiujiang Guo}
\affiliation{School of Physics, and ZJU-Hangzhou Global Scientific and Technological Innovation Center, Zhejiang University, Hangzhou 310027, China}
\author{Junxiang Zhang}
\affiliation{School of Physics, and ZJU-Hangzhou Global Scientific and Technological Innovation Center, Zhejiang University, Hangzhou 310027, China}

\author{Lei Ying}\email{leiying@zju.edu.cn}
\affiliation{School of Physics, and ZJU-Hangzhou Global Scientific and Technological Innovation Center, Zhejiang University, Hangzhou 310027, China}

\author{Zlatko Papi\'c}\email{z.papic@leeds.ac.uk}
\affiliation{School of Physics and Astronomy, University of Leeds, Leeds, UK}

\begin{abstract}
Emerging quantum technologies hold the promise of unraveling difficult problems ranging from condensed matter to high energy physics, while at the same time motivating the search for unprecedented phenomena in their setting. Here we utilize a custom-built superconducting qubit ladder to realize non-thermalizing states with rich entanglement structures in the middle of the energy spectrum. Despite effectively forming an “infinite” temperature ensemble, these states robustly encode quantum information far from equilibrium, as we demonstrate by measuring the fidelity and entanglement entropy in the quench dynamics of the ladder. Our approach harnesses the recently proposed type of non-ergodic behavior known as “rainbow scar”, which allows us to obtain analytically exact eigenfunctions whose ergodicity-breaking properties can be conveniently controlled by randomizing the couplings of the model, without affecting their energy. The on-demand tunability of quantum correlations via disorder allows for \emph{in situ} control over ergodicity breaking and it provides a knob for designing exotic many-body states that defy thermalization.
\end{abstract}

\date{\today}
\maketitle

\section*{Introduction}\label{sec:intro}
The abundance of entanglement and other types of correlations in many-body systems make them an attractive resource for quantum information processing. Quantum coherence, however, is typically fragile, even in systems that can be considered well-isolated from the environment: coherence rapidly deteriorates in the presence of a finite density of quasiparticle excitations above the systems' ground state. This is a consequence of thermalization -- the ultimate fate of generic systems comprising many interacting degrees of freedom~\cite{DeutschETH, SrednickiETH, RigolNature}.   If thermalization breaks down,  new types of dynamical behavior and phases of matter can emerge.  For example, finely tuned one-dimensional systems~\cite{Kinoshita06} can evade thermalization due to their rich symmetry structure known as quantum integrability~\cite{Sutherland}. On the other hand, in real materials, disorder is ubiquitous and, if strong enough, it can strongly suppress thermalization by turning the system into an Anderson insulator~\cite{Anderson58} or its interacting cousin, the many-body localized (MBL) phase~\cite{Huse-rev, AbaninRev}. 

The ability to suppress thermalization while retaining a high degree of control over entanglement is key to robust technological applications based on many-body quantum systems. Integrable systems do not meet this requirement as they are restricted to one spatial dimension and require fine tuning of the parameters. In MBL systems, bulk excitations are localized, regardless of their energy density, which indeed can effectively protect the information stored in the degrees of freedom at the system's boundary~\cite{Huse13L,Bahri:2015aa}. Nevertheless, entanglement in MBL states is typically bounded by ``area-law'' scaling~\cite{Bauer13}. This limits their applications in quantum-enhanced metrology, which often rely on large multipartite entanglement~\cite{PezzeRMP}. The latter entanglement structures have recently been identified~\cite{DesaulesQFI,Dooley2023} in a class of systems known as quantum many-body scars (QMBS)~\cite{Serbyn2021, MoudgalyaReview, ChandranReview}.
When QMBS systems are prepared in special initial states, their dynamics become trapped in a subspace that does not mix with the thermalizing bulk of the spectrum, leading to the coherent time evolution of local observables~\cite{Bernien2017, Bluvstein2021, Jepsen2021, GuoXian2022, zhang2022many,yao2022observation}.
The observation of QMBS has triggered a flurry of theoretical efforts to understand and classify the general mechanisms of weak ergodicity breaking in isolated quantum systems~\cite{ShiraishiMori, BernevigEnt, Turner2018a, wenwei18TDVPscar, Iadecola2019_2, MotrunichTowers, Dea2020, Pakrouski2020, Moudgalya2022, Buca2023}. 

In this work, inspired by our state-of-the-art superconducting qubit processor in which the qubit-qubit coupling can be broadly tuned to encompass opposite coupling signs, we demonstrate the existence of entanglement structures that persist far from equilibrium and can be \emph{deterministically} tuned by disorder. The approach is inspired by the rainbow scar construction~\cite{Langlett2022,Wildeboer22}, which creates Bell pairs between qubits belonging to two halves of the system. We show that our model hosts several distinct families of QMBS states and entanglement structures. While the first family is a direct realization of the rainbow construction, the second family emerges from a hitherto unexplored mechanism: it is obtained by acting on the first family with the Hamiltonian of a \emph{single} subsystem. By making the couplings spatially inhomogeneous, we can then turn these states into \emph{disordered} QMBS states whose \emph{exact} wave functions and entanglement structure can still be written down in analytic form.  Unlike their energies, the structure of these exact eigenstates can be explicitly modulated via the disorder profile, allowing tuning of their properties. We experimentally observe the two types of entanglement via their characteristic ergodicity-breaking signatures in quantum dynamics at late times. 

\section*{Results} \label{sec:results}

\begin{figure}
\centering
\includegraphics[width=\linewidth]{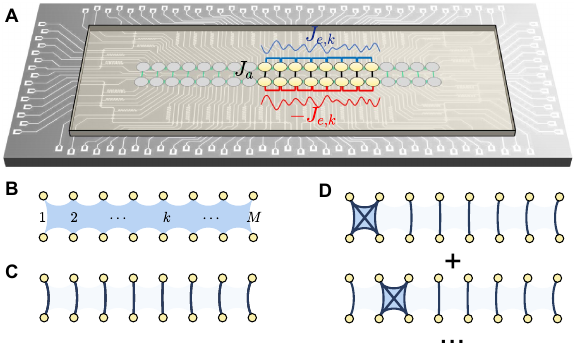}
\caption{
{\bf The device and entanglement structures.}
({\bf A}) Micrograph of the superconducting quantum processor in a ladder configuration. The tunable couplings, $J_a$ and $J_e$, between nearest-neighbor qubits, belonging to the same or opposite rows, are indicated. Blue and red curves illustrate the disordered coupling strengths $\pm J_{e,k}$, carrying opposite signs in the two rows. ({\bf B}-{\bf D}) Schematic representation of entanglement structure for a thermalizing state, the first family, and second family of scars, respectively. The shaded blue region in ({\bf B}) indicates large entanglement between all qubits in the thermalizing case. In ({\bf C}-{\bf D}), dark curves depict the Bell pair entanglement of neighboring qubits, with dimers locally forming $|\mathbf{T}\rangle$ states. The tetramer configurations denote doublon-holon entanglement  $\left(|{\mathbf{DH}}\rangle+|\mathbf{HD}\rangle\right)/\sqrt{2}$, characteristic of the second scar family. 
}
\label{fig_schematic}
\end{figure}

{\bf The model and its symmetries.---}Our 
superconducting quantum processor contains $N=2M$ qubits arranged in a ladder configuration, depicted in Fig.~\ref{fig_schematic}{\bf A}, with two horizontal rungs containing $M$ qubits each. The coupling strength between a pair of nearest-neighbor transmon qubits can be capacitively tuned by a coupler, enabling broad ranges of $[-8,8]$~MHz and $[-8,-2]$~MHz for the parallel and vertical couplings, respectively, see Supplementary Material (SM) for further characteristics of the device~\cite{SOM}.
We denote the states of each qubit by $\ket{\circ}$ and $\ket{\bullet}$. The qubits belonging to the top row are described by $\hat{\mathbf{u}}^\alpha$ Pauli matrices (with $\alpha=x,y,z$), while $\hat{\mathbf{d}}^\alpha$ are Pauli matrices acting on the bottom-row qubits. The Hamiltonian can be written as
\begin{equation}\label{eq:model}
    \hat{H}=\hat{H}_{\mathbf{u}}\otimes \mathbb{1} +\mathbb{1}\otimes \hat{H}_\mathbf{d}+\hat{H}_\mathrm{int},
\end{equation}
where the top/bottom row and inter-row Hamiltonians, respectively, are given by
\begin{eqnarray}\label{eq:ourH1}
 \notag  \hat{H}_{\sigma=\mathbf{u},\mathbf{d}}/2\pi &=&  \sum_{k=1}^{M-1}  \pm \frac{ J_{e,k}}{2}\left( \hat \sigma^x_k \hat \sigma^x_{k+1}{+} \hat \sigma^y_k \hat \sigma^y_{k+1} \right)  
    +\sum_{k=1}^M \omega_k \hat \sigma^z_k, \\
    \hat{H}_\mathrm{int}/2\pi &=& \ \ \ \sum_{k=1}^M \frac{J_a}{2} \left( \hat{\mathbf{u}}^x_k\hat{\mathbf{d}}^x_k + \hat{\mathbf{u}}^y_k\hat{\mathbf{d}}^y_k \right).   
\end{eqnarray}
Here, the intralayer coupling amplitude $J_{e,k}$ and the frequency $\omega_k$ can be site-dependent, allowing for the possibility of disorder, while the inter-row coupling $J_{a}$ is required to be uniform~\cite{SOM}. The rainbow construction mandates that the bottom row of qubits must have intra-row coupling amplitudes of the opposite sign. One can easily verify that the two Hamiltonians are related by the mirror transformation 
$\hat{H}_\mathbf{d} = -\mathcal{M}\hat{H}_\mathbf{u}^\star\mathcal{M}^\dagger$, where $\cal M$ simply maps $\ket{{\circ}}_{\mathrm{u},k} \leftrightarrow\ket{{\bullet}}_{\mathrm{d},k}$ and $\ket{{\bullet}}_{\mathrm{u},k} \leftrightarrow\ket{{\circ}}_{\mathrm{d},k}$~\cite{SOM}. Crucially, the mirror transformation forces the spectra of $\hat{H}_\mathbf{u}$ and $\hat{H}_\mathbf{d}$ to be identical but of opposite signs. 

The model in Eq.~(\ref{eq:model}) possesses a U(1) symmetry corresponding to the conservation of total magnetization along the $z$-direction and below, unless specified otherwise, we will restrict to its largest sector with zero magnetization or half-filling. Interestingly, the exchange rules of the Hamiltonian give rise to an additional, more subtle, symmetry. 
Our Hamiltonian can exchange neighboring triplet states, $\mathbf{T} \equiv \left({{\circ \atop \bullet}}+{{\bullet \atop \circ}}\right)/\sqrt{2} $ and singlet states, $\mathbf{S} \equiv \left({{\circ \atop \bullet}}-{{\bullet \atop \circ}}\right)/\sqrt{2}$. Alternatively, it can create a domain with a doublon $\mathbf{D}\equiv \left({{\bullet \atop \bullet}}\right)$ or holon $\mathbf{H}\equiv \left({{\circ \atop \circ}}\right)$ on each side in which the $\mathbf{T}$ and $\mathbf{S} $ are exchanged. Thus, $\hat H$ conserves the difference between the number of triplets and singlets, 
multiplied by a phase factor that counts the number of doublons and holons to the left of a given site (see Methods for a formal definition of $\hat Q$). The $\hat{Q}$ symmetry further splits the half-filling sector of the Hilbert space into $M+1$ disconnected sectors with quantum numbers $-M,-M+2,\ldots,M-2,M$.  Working in the largest $Q$ sector (at zero magnetization), we checked the statistics of the energy level spacings of $\hat H$ using exact diagonalization, finding excellent agreement with the Wigner-Dyson ensemble and very small fluctuations between different disorder realizations (see Methods), thus indicating that the model is quantum-chaotic. 

\begin{figure}
\centering
\includegraphics[width=\linewidth]{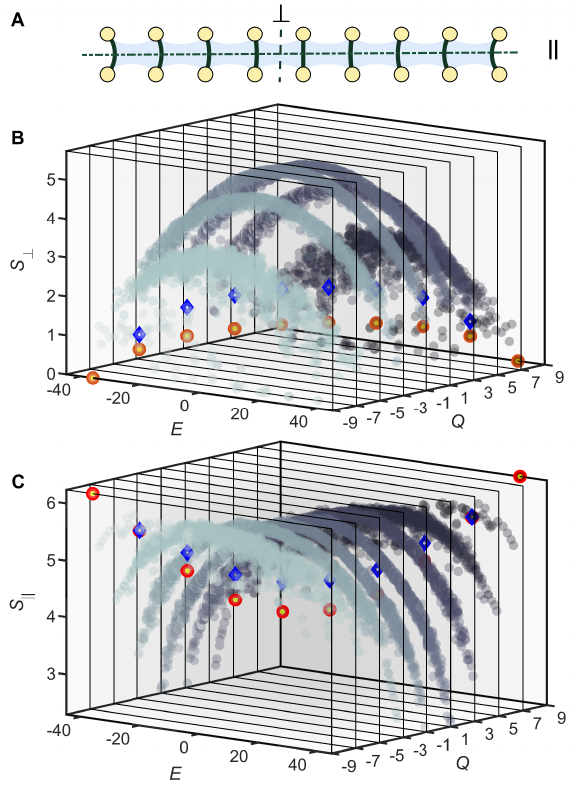}
\caption{
{\bf Rainbow entanglement.} ({\bf A}) Schematic of the ladder with dashed lines indicating two types of bipartitions.  The parallel bipartition splits all entangled pairs in the rainbow state, while the perpendicular bipartition does not split any pair. ({\bf B}-{\bf C}) Bipartite entropy of eigenstates in different $Q$-symmetry sectors for the two cuts shown in ({\bf A}). The two families of scarred states are highlighted by red circles (first family) and blue diamonds (second family). Their rainbow nature is revealed by the fact that they move between minimal and maximal entropy, depending on the cut.  The upper bounds of bipartite entropy in ({\bf B}) and ({\bf C}) are given by the Page entropy $(2M\ln 2-1)/2$~\cite{Page93} and the maximum subsystem entropy $M\ln2$, respectively. Colored cross sections represent different sectors labeled by the values on the $Q$ axis. Data is obtained by exact diagonalization with $N=18$ qubits, $J_a=4$, and $J_{e,k}\in [4,4.5]$, $\omega_k\in [0.5,1.5]$ drawn from a uniform distribution. 
}
\label{fig:model}
\end{figure}

{\bf Two families of rainbow scar entanglements.---}The rainbow state $\ket{\mathbf{I}}=\ket{\mathbf{T}\mathbf{T}\ldots \mathbf{T}}$ is an eigenstate of our model in Eq.~(\ref{eq:model}). Two different families of scars can be built from it using operators that commute with the half-system Hamiltonian $\hat H_\mathbf{u}$.
To construct the first family, we use the operator $\hat{Z}=\sum_{k=1}^M \hat{\mathbf{u}}^z_k$, which clearly commutes with $\hat H_\mathbf{u}$ as the latter conserves $z$-magnetization. The powers of $\hat{Z}$ are linearly independent up to $\hat{Z}^M$, thus we can apply $\hat{Z}$ up to $M$ times. The $\hat Z$ operator simply converts triplets into singlets and vice versa. The resulting states $\ket{M-n,n}
$ will be symmetric superpositions of all states with a fixed number $n$ of triplets and $M{-}n$ singlets. The scarred states of the first family are precisely these states, up to normalization:
\begin{equation} \label{eq:E_scar}
    \ket{E_n}=\binom{M}{n}^{-1/2}\ket{M{-}n,n}\propto \hat{P}^Q_{2n-M}\hat{Z}\ket{E_{n+1}},
\end{equation}
where $n$ ranges between $0$ and $M-1$ and $\ket{E_M}\equiv \ket{\mathbf{I}}$. The second equality illustrates that we can build $\ket{E_n}$ recursively from $\ket{E_{n+1}}$ by making use of the projector $\hat{P}^Q_q$ on the sector of $\hat{Q}$ with an eigenvalue $q$. The projector $\hat{P}^Q_q$ is introduced for convenience as it simplifies the recursion, since acting with $\hat{Z}$ on $\ket{E_n}$ creates a superposition of $\ket{E_{n-1}}$ and $\ket{E_{n+1}}$. It can be verified that the state $\ket{E_n}$ is an eigenstate of $\hat{H}$ with energy $E_n=J_a (2n-M)$~\cite{SOM}. 

While symmetry generators come to mind when we look for operators $\hat O$ that commute with $\hat{H}_\mathbf{u}$, we can build a different family of scarred states by using $\hat{H}_\mathbf{u}$ itself as the generator. This gives us the second family of scars:
\begin{equation}\label{eq:secondtype}
    \ket{E^\prime_n}{\propto}\hat{P}^Q_{M{-}2n}\left[\hat{H}_\mathbf{u}{-}\left(\sum_k \frac{\omega_k}{M}\right)\hat{Z}\right]\ket{E_{n+1}},
\end{equation}
with $n=1,2,\ldots,M-1$. The second term in the square bracket automatically orthogonalizes the states $\ket{E_n^\prime}$ with respect to the first family of scars. The projectors $\hat{P}^Q_q$ once again isolate $\ket{E^\prime_{n-1}}$ from $\ket{E^\prime_{n+1}}$. Similar to the first type of scar, the second type of scarred states are also equidistant in energy, in fact occurring at the same energies $E_n=J_a (2n-M)$~\cite{SOM}.

It is instructive to contrast the two families of scars, Eq.~(\ref{eq:E_scar}) and (\ref{eq:secondtype}). While both families occur at the same, regularly spaced energies throughout the spectrum, the total number of states in the second family is smaller by two than in the first family. Furthermore, there are stark differences in entanglement structures. The states belonging to the second family explicitly depend on disorder through their dependence on $J_{e,k}$ and $\omega_k$, unlike the first family. The states in the first family contain only singlets or triplets, with no doublons or holons, Fig.~\ref{fig_schematic}{\bf C}. By contrast, the second family has overlap with states involving a symmetric superposition on a single doublon-holon pair  $\ket{{\ldots} \mathbf{DH}{\ldots}}+\ket{{\ldots} \mathbf{HD}{\ldots}}$ with weight $J_{e,k}/2$ on top of a background of $M{-}n{-}1$ singlets and $n{-}1$ triplets, Fig.~\ref{fig_schematic}{\bf D}. They also have overlap with all states with $M{-}n$ singlets and $n$ triplets, with prefactors depending on the $\omega_k$ and on the location of the triplets. The dependence of $\ket{E^\prime_n}$ states on $\omega_k$ and $J_{e,k}$ allows us to tune their properties, such as entanglement entropy or overlap with special initial states. 

The two families of scars are further identified by their entanglement entropy, ${S}_{A}=-\mathrm{tr} \rho_{A} \log \rho_{A}$, where $\rho_{A}$ is the reduced density matrix of the subsystem $A$. The reduced density matrix $\rho_A = \mathrm{tr}_{\bar A}\rho$ is obtained from the full density matrix $\rho$ by tracing out the degrees of freedom belonging to the complement $\bar A$ of the subsystem $A$. The rainbow entanglement manifests as a striking difference in entropy depending on the type of bipartition that defines the subsystem $A$, and we will consider two types illustrated in Fig.~\ref{fig:model}{\bf A}. For the parallel cut between the two rows of the ladder, i.e., when $A$ comprises qubits $ \{u_1,u_2,\ldots, u_M \}$, the entanglement is large, as the bipartition cuts through an extensive number of Bell pairs. By contrast,  for the bipartition perpendicular to the ladder, i.e., when $ A = \{u_1,d_1,\ldots, u_{M/2},d_{M/2} \}\equiv\{k=1,2,\cdots,M/2\}$,  the entanglement is much lower. As seen in Fig.~\ref{fig:model}{\bf B}-{\bf C}, this distinction is particularly striking for $\ket{E_0}$ and $\ket{E_M}$ states, which have nearly \emph{maximal} entropy (i.e., scaling with the number of qubits) in the case of parallel bipartition but low entanglement (i.e., bounded by a constant) in the case of a perpendicular bipartition. Furthermore, for the first type of scarred eigenstate in the middle of the spectrum ($E{=}0$), it can be analytically shown  that the maximum bipartite entanglement is $S_{1,\perp}^{M\to\infty} = \left[1+\ln(\pi M/8)\right]/2$, 
while for the second type of scarred eigenstates, we numerically established that the disorder allows us to tune the entanglement in the range $(S_{1,\perp}, S_{1,\perp} + \ln 4)$~\cite{SOM}.

\begin{figure}[tb]
\centering
\includegraphics[width=\linewidth]{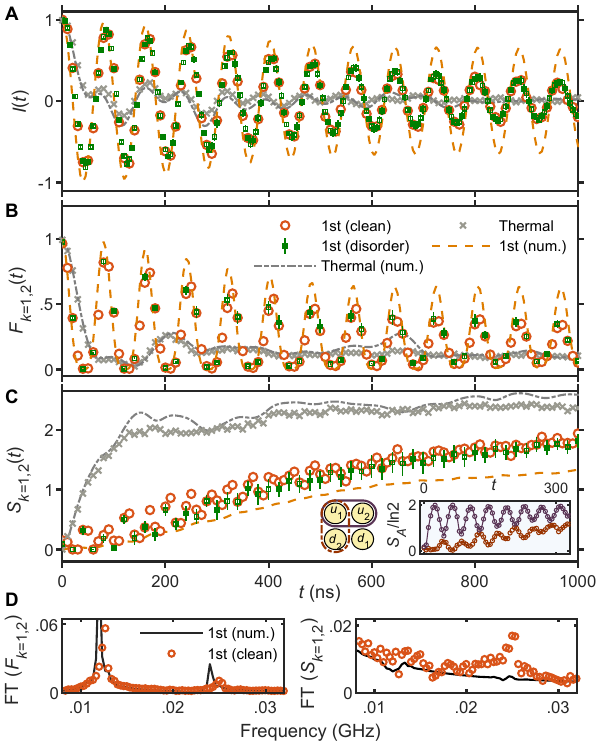}
\caption{
{\bf Experimental observation of the dynamical signatures of the first scar family.} The ladder with $N=10$ qubits is initialized in the state $\ket{\Pi}$ and the couplings are set to values $J_{e,k}=J_e=J_a/3= -2$~MHz. The plots show the measurements of ({\bf A}) population imbalance, ({\bf B}) the $4$-qubit fidelity, and ({\bf C}) the $4$-qubit entanglement entropy, specified in the text. The inset in ({\bf C}) shows the entropy of a subsystem consisting of 2 qubits, sketched on the left. Purple dots with error bars stand for the average and standard deviation over 8 disorder realizations of $J_{e,k}$, randomly selected from an interval $\in[1,3]$~MHz. For reference, we also show a typical initial state that thermalizes (see text for details). The lines are the results of numerical simulations for the same parameters, including the additional cross couplings $J_x \approx 0.3$~MHz and nonlinearity of qubits $\eta\approx -175$~MHz, present in the physical device (see Methods).  The subscripts $k=1,2$ on the fidelity and entanglement denote the measured subsystem. ({\bf D}) Fourier spectrum of fidelity and entropy dynamics in panels {(\bf B}) and ({\bf C}), respectively.
}
\label{fig:exp_scar1}
\end{figure}

{\bf Entanglement dynamics.---}To observe rainbow entanglements, we utilize an established diagnostics of QMBS~\cite{Bernien2017}: the evolution of local observable expectation values in quench dynamics of the circuit, consisting of two contiguous rows with up to 8 qubits each.  The structure of the first family of scarred eigenstates implies that they have high overlap with the product state $\ket{\Pi}{=}\ket{{\bullet\atop\circ}{\bullet\atop\circ}\cdots {\bullet\atop\circ}}$. Fig.~\ref{fig:exp_scar1}{\bf A} shows the dynamics of population imbalance, ${I}(t) =  ({1}/{N})  \sum_{k=1}^M \sum_{\sigma=\mathbf{u},\mathbf{d}}  \langle \hat{\sigma}^z_k(0)  \rangle   \langle\hat{\sigma}^z_k(t) \rangle$. The population imbalance in the $\ket{\Pi}$ state exhibits remarkable oscillations that persist up to time scales ${\sim} 1\mu$s. This is in contrast with a typical thermalizing state  $\ket{{\bullet\atop\bullet}{\circ\atop\circ}{\bullet\atop\bullet}{\circ\atop\circ}{\bullet\atop\circ}}$, for which the population imbalance rapidly decays to zero by ${\sim} 50$\;ns. A salient feature of the first family of scars is their insensitivity to the disorder in the Hamiltonian couplings. Thus, we expect the coherent dynamics from the $\ket{\Pi}$ state to be unchanged when inhomogeneity is introduced in $J_{e,k}$ couplings. This signature is clearly confirmed by experimental observations in Fig.~\ref{fig:exp_scar1}{\bf A}.

Furthermore, we utilized the quantum tomography technique and obtained the reduced density matrix of the subsystem consisting of qubits $A=\{k=1,2\}$, which gives us additional information about the dynamics beyond local observables.  The subsystem fidelity, $\sqrt{F_{A}} = \mathrm{tr}\sqrt{\sqrt{\rho_{A}(t)}\rho_{A}(0)\sqrt{\rho_{A}(t)}}$, and entanglement entropy, ${S}_{k=1,2}$, are shown in Figs.~\ref{fig:exp_scar1}{\bf B}-{\bf C}. For the initial state $\ket{\Pi}$, the subsystem fidelity dynamics undergoes persistent revivals, implying that the initial information is restored many times, with a period of about $80$~ns. Meanwhile, for generic initial product states, ${F}_{k=1,2}$ quickly decays towards a value close to the inverse of the subsystem Hilbert space dimension, as shown in Fig.~\ref{fig:exp_scar1}{\bf B}. The growth of entanglement entropy also shows a stark contrast between initial states.
Compared to the thermal states, the $\ket{\Pi}$ state exhibits a slow linear growth with superposed oscillations. The small oscillations are in correspondence with the peaks and valleys observed in the fidelity dynamics, with roughly half the period of the latter, as shown as the Fourier spectrum of the fidelity and entropy dynamics in Fig.~\ref{fig:exp_scar1}{\bf D}.
Furthermore, we show the entropy dynamics with different subsystems $\{u_1,u_2\}$ and $\{u_1,d_1\}$ in the inset of Fig.~\ref{fig:exp_scar1}{\bf C}, confirming the rainbow entanglement structure previously sketched in Fig.~\ref{fig_schematic}{\bf C}. 

We note that the $\ket{\Pi}$ initial state can be proven to exhibit \emph{perfect} revivals and constant-in-time entanglement entropy for our model in Eq.~(\ref{eq:model})~\cite{SOM}. In contrast, Fig.~\ref{fig:exp_scar1} shows a weak population and fidelity decay, along with a slow growth of entanglement. Detailed characterization of the experimental device revealed two extraneous terms not present in the theoretical model, which correspond to diagonal XY couplings and to three-photon occupation not being fully suppressed~\cite{SOM}. Numerical simulations, shown by lines in Fig.~\ref{fig:exp_scar1},  confirm that these perturbations capture the main sources of decay of local observables and entanglement growth (see Methods). The effect of these perturbations, however, is sufficiently weak such that clear signatures of the two families of scars can be observed and sharply distinguished, as shown next.

\begin{figure}[tb]
\centering
\includegraphics[width=\linewidth]{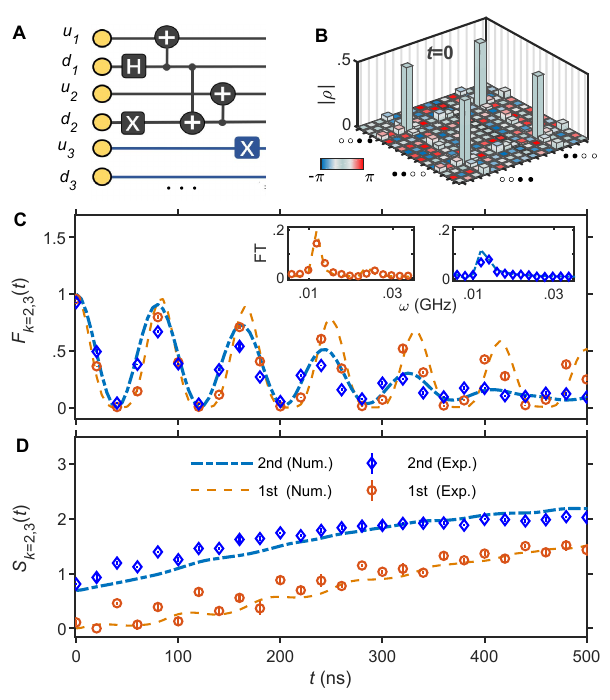}
\caption{
{\bf Experimental distinction between the first and second family of scars.}  
({\bf A}) Circuit diagram for generating the entangled initial state $\ket{\phi_L}$  used to probe the dynamics of the second scar family. Symbols ``$+$'',``$\mathrm{H}$'', and ``$\mathrm{X}$'' stand for CNOT, Hadamard, and Pauli-X gates, respectively.
({\bf B}) Absolute values of the reduced density matrix elements  $\rho_{k=1,2}$ at $t=0$, with the color bar denoting their phase. ({\bf C}-{\bf D}) Fidelity and entanglement entropy dynamics of a 4-qubit subsystem for initial states $|\Pi\rangle$ and $|\phi_L\rangle$, which overlap with the first and second family of scars, respectively. Insets of ({\bf C}) show the Fourier spectra of the fidelity, which distinguish the second family (only one peak) from the first family (two peaks).
The superconducting ladder contains $N=16$ qubits with the same parameters as Fig.~3.
}
\label{fig:exp_scar3}
\end{figure}

\begin{figure}[tb]
\centering
\includegraphics[width=\linewidth]{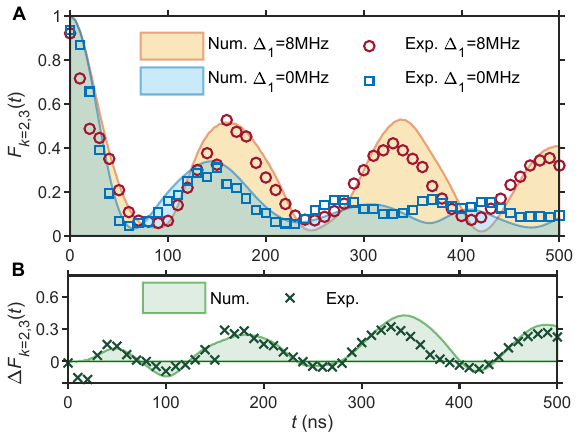}
\caption{
{\bf Disorder-tunable scar of the second family.} ({\bf A}) Fidelity $F_{k=2,3}$ for a tunability of $\Delta_1=0$ (blue square) or $\Delta_1=8$~MHz (red circle) over the first inter-dimer coupling $J_{e,1} = J_e+\Delta_1$. The experimental data (markers) is for $N=8$ qubits, $J_a=3.0$~MHz, $J_{e,k}\in [2.0,3.0]$~MHz, drawn from a uniform distribution, while the curves are from the numerical simulations based on the experimental model. ({\bf B}) The fidelity difference $\Delta F_{k=2,3}$ between the two cases in ({\bf A}), illustrating the revival enhancement by $\Delta_1$. Regions with negative $\Delta F_{k=2,3}$ are due to a slight shift of the revival peaks. 
}
\label{fig:exp_scar4}
\end{figure}

To probe the second scar family,  we require  a more complicated initial state with an entangled doublon-holon pair:
$\ket{\phi_L}{=}\frac{1}{\sqrt{2}}\Big(\ket{{\bullet \atop \bullet}{\circ \atop \circ}}{+}\ket{{\circ \atop \circ}{\bullet \atop \bullet}}\Big)\otimes\ket{{\circ \atop \bullet}\cdots {\circ \atop \bullet}}$, which has predominant overlap with the second family of scarred eigenstates~\cite{SOM}. In contrast with the $\ket{\Pi}$ state, the state $\ket{\phi_L}$ also has a small overlap on the non-scarred subspace, thus we expect slowly decaying revivals in the latter case, even in the absence of any perturbations. We emphasize that the state $\ket{\phi_L}$ is orthogonal to the first family of scars, as the latter do not contain any doublons or holons. To prepare the state $\ket{\phi_L}$, we use the circuit scheme in Fig.~\ref{fig:exp_scar3}{\bf A}, which is composed of a few single-qubit and two-qubit gates. By utilizing high-precision tomography measurements, we then obtain the reduced density matrix of the subsystems $\{k=1,2\}$ or $\{k=2,3\}$. The former one at $t=0$ is visualized in Fig.~\ref{fig:exp_scar3}{\bf B}, demonstrating that the entangled state $\ket{\phi_L}$ is successfully prepared. 

To reveal the difference between the first and second family of scars, we focus on the subsystem $A^\prime\equiv\{k=2,3\}$, whose fidelity $F_{k=2,3}(t)$ and entanglement entropy $S_{k=2,3}(t)$, are plotted in Figs.~\ref{fig:exp_scar3}{\bf C}-{\bf D}. 
The subsystem fidelity partially reveals the scarred eigenstates and the Fourier transformation of $F_{k=2,3}$ for the state $\ket{\Pi}$ has an additional peak compared to the state $\ket{\phi_L}$. 
This difference is related to the fact that the first family of scarred eigenstates contains two more members compared to the second family in Fig.~\ref{fig:model}{\bf B}. Since only four qubits $\{k=2,3\}$ are observed and the coherent information of the rest of the system is traced out, the dynamics of $F_{k=2,3}(t)$ approximately reflects the subsystem itself. Thus, the Fourier spectrum of $F_{k=2,3}(t)$ involves two peaks and one peak for state $\Pi$ and $\phi_L$ due to only two and one dimers, respectively~\cite{SOM}.
Furthermore, the choice of the subsystem is motivated by the fact that it leads to entropy $\ln 2$ for the $\ket{\phi_L}$ initial state, while the entropy is still trivially zero for the $\ket{\Pi}$ state. This distinction is verified in our experiment, as shown in Fig.~\ref{fig:exp_scar3}{\bf D}. 

{\bf Tunable revivals of the second scar family.---}The revivals associated with the second scar family can be conveniently tuned by modulating individual couplings $J_{e,k}$, even in the presence of experimental imperfections. The underlying mechanism behind the revival tunability can be understood by considering the projection of $\ket{\phi_L}$  state on the set of scarred eigenstates  $\ket{E^\prime_n}$, which can be shown to be~\cite{SOM}
\begin{equation}\label{eq:olap_L_tot}
    \sum_{n=1}^{M-1}|\langle \phi_L | E_n^\prime \rangle|^2=\frac{J_{e,1}^2}{ \sum_{k=1}^{M-1}J_{e,k}^2{+}2\sum_{k=1}^{M}\omg_k^2  },
\end{equation}
with $\omg_k \equiv \omega_k - \sum_j \omega_j/M$. Thus, within the model~(\ref{eq:model}), we recover perfect revivals in the limit $J_{e,1}{\to}\infty$. 
To illustrate the $J_{e,1}$ tunability, in Fig.~\ref{fig:exp_scar4} we measure the fidelity dynamics as $J_{e,1}$ is modulated by an amount $\Delta_1 \in [0,8]$. Within the accessible range of $\Delta_1$, the model remains chaotic yet we observe an increase in fidelity of about 0.3, consistent with the theoretical prediction. This demonstrates that the tunability of the revivals for the second scar family can be achieved with the experimentally relevant values of the parameters.

\section*{Discussion} 
We have realized multiple families of non-thermalizing states with distinct rainbow entanglement structures throughout the energy spectrum. Our construction allows to write down exact wave functions for these states, even in the presence of disorder.  The existence and stability of scar states in disordered models have recently attracted much attention in theoretical studies~\cite{OnsagerScars,MondragonShem2020,Huang2021, Voorden21, Zhang2022Clusters}. 
A unique aspect of our model is that it allows to tailor the entanglement of scarred states, while the explicit eigenstate dependence on the disorder profile controls the extent of ergodicity breaking. Signatures of rainbow entanglements were observed by performing quantum state tomography of a many-body state of the ladder following the quench from special initial states, confirming the expected hallmarks of QMBS behavior, such as robust revivals in the fidelity dynamics and slow growth of entanglement far from equilibrium. 

While in this work the disorder strength was assumed to be sufficiently weak such that the system overall remains chaotic, the versatility of our setup allows direct access to \emph{strong} ergodicity-breaking regimes, where many-body localization was recently proposed to give rise to ``inverted scarring'' phenomena~\cite{Srivatsa2022,Chen23,Iversen23}. More generally, our work bridges the gap between theoretical studies of QMBS, which place the emphasis on exact constructions of scarred eigenstates~\cite{MoudgalyaReview, ChandranReview}, and experimental realizations, e.g., in Rydberg atom arrays~\cite{Bernien2017,Bluvstein2021} or optical lattices~\cite{GuoXian2022}, in which the scarred states  are not known exactly (apart from a few exceptions~\cite{lin2018exact}). In contrast, our model (\ref{eq:model}) hosts exact scars, while its experimental implementation contains additional perturbations. While these perturbations were shown to be  sufficiently weak in our device to allow unambiguous observation of scar signatures even in classical simulations, it would be interesting to study in detail their effects on the stability of QMBS states in larger circuits or higher-dimensional geometries that would rapidly exceed the capability of classical computers.

The flexibility of our construction stems from the fact that the scarred states are not generated by conventional symmetry action, but by the Hamiltonian describing one of the rungs of the ladder. For simplicity, we assumed the latter describes an integrable XY model (although the system overall is non-integrable). This was not essential, however, and our construction can be straightforwardly generalized to cases where the subsystem Hamiltonian is non-integrable~\cite{SOM}. The key to implementing the construction was the broad tunability of the experimental device that allowed us to vary the coupling sign, in contrast with traditional multi-qubit superconducting systems~\cite{AABR:2019}. This tunability offers an additional physical freedom that can be used for designing exotic many-body states that defy quantum thermalization and information scrambling, in particular the creation of multipartite-entangled states without the need for larger spins or complicated interactions~\cite{Dooley2023}.

\section*{Materials and Methods}

\begin{figure}[tb]
\centering
\includegraphics[width=\linewidth]{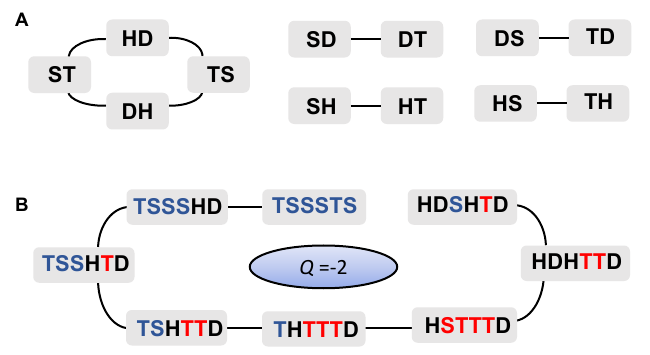}
\caption{
{\bf Hamiltonian action and conservation laws.}
({\bf A}) Schematic of the Hamiltonian action on neighboring dimers.
({\bf B}) Example of a sequence of states obtained using only allowed processes. The color of the singlet and triplet states indicates if they have been ``exchanged" an even (blue) or odd (red) number of times. This exactly corresponds to the parity of the sum of the number of holons and doublons to their left. If one gets $+1$ for all blue triplets and red singlets, and $-1$ for all blue singles and red triplets, then for all states in the sequence the sum is the same and gives $Q{=}{-}2$, illustrating the conservation of this charge.
}
\label{fig:ladder_sym}
\end{figure}

{\bf Symmetries of the model and statistics of energy levels.---}
Our model, defined in Eq.~(\ref{eq:model}), conserves the following quantity $\hat Q$:
\begin{equation}\label{eq:Q}
    \hat{Q}=\sum_{k=1}^M \left(\hat{T}_k-\hat{S}_k\right)\prod_{l=1}^{k-1}(-1)^{\hat{H}_l+\hat{D}_l},
\end{equation}
where $\hat{T}_k$, $\hat{S}_k$, $\hat{D}_k$ and $\hat{H}_k$ are projectors on the respective dimer state at the given site $k$. Here we outline the proof of this statement and provide physical intuition behind the conservation of $\hat Q$.  We start by considering the action of the Hamiltonian on the dimer basis introduced in the main text. The relevant dynamical terms are summarized in Fig.~\ref{fig:ladder_sym}{\bf A}. Consider a simple configuration like $\mathbf{TSSSTS}$. The Hamiltonian rules dictate that we can only exchange $\mathbf{TS}$ or $\mathbf{ST}$ into $\mathbf{HD}$ or $\mathbf{DH}$. Therefore, a natural guess for the conserved quantity would be the difference between the total number of triplets and the number of singlets: $\sum_k(\hat{T}_k-\hat{S}_k)$. This works perfectly until we start having configurations with neighboring dimers like $\mathbf{SH}$. In that case, we can turn it into $\mathbf{HT}$ and thus change $\sum_k(\hat{T}_k-\hat{S}_k)$. However, for that to happen it means that we first created $\mathbf{HD}$ and that the dimer that was switched between $\mathbf{S}$ and $\mathbf{T}$ must be surrounded by  $\mathbf{H}$ and $\mathbf{D}$. Effectively, the Hamiltonian creates domains inside which all $\mathbf{T}$ and $\mathbf{S}$ are ``exchanged''. Thus, we can keep track of how many times such exchanges have occurred for a given dimer by counting the number of doublons or holons to its left. Fig.~\ref{fig:ladder_sym}{\bf B} shows an example of this process.

\begin{figure}[tb]
\centering
\includegraphics[width=0.9\linewidth]{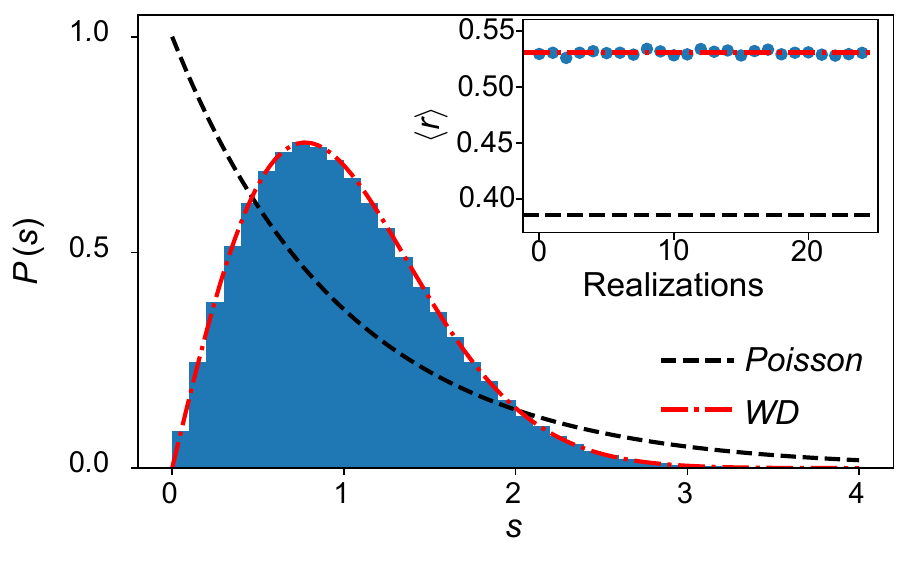}
\caption{{\bf Level statistics.}  Distribution of energy level spacings $s\equiv E_{n+1}-E_n$ for the model in Eq.~(\ref{eq:model}) with $N{=}20$ sites. Data is for half-filling and $Q{=}0$, with 25 disorder realizations. The level statistics displays excellent agreement with the Wigner-Dyson ensemble. The inset shows the average consecutive spacing ratio $\langle r \rangle$ for each realization, which is very close to the expected value of 0.53 for a chaotic system. Data is for $J_a{=}3$ and $J_{e,k}\in [2,2.5]$, $\omega_k\in [0.5,1.5]$ drawn from a uniform distribution.
}
\label{fig:level_stats}
\end{figure}

While we focused on the half-filling sector in the main text, we note that $\hat{Q}$ is a symmetry at any filling, see SM for a detailed discussion~\cite{SOM}. The interplay of $\hat Q$ and filling leads to a large number of disconnected sectors. Some of them are of small dimension and similar to the ones studied in Ref.~\cite{IadecolaLocalization}. Nonetheless, for large sectors we find good agreement with the eigenstate thermalization hypothesis (ETH) predictions, as shown in Fig.~\ref{fig:level_stats}. 

Finally, beyond the symmetry $\hat{Q}$, at half-filling the Hamiltonian \emph{anticommutes} with 
\begin{equation}
\hat{C}=\hat{\mathcal{P}}_{\mathbf{d}\leftrightarrow \mathbf{u}}\prod_{k=1}^M\hat{\mathbf{u}}^x_k\hat{\mathbf{d}}^x_k\hat{\mathbf{d}}^z_k
\end{equation}
where $\hat{\mathcal{P}}_{\mathbf{d}\leftrightarrow \mathbf{u}}$ is an operator exchanging the top and bottom rows. $\hat{C}$ is related to a particle-hole transformation along with a swap between the top and bottom row and an additional phase. In the dimer picture, $\hat{C}$  simply switches $\mathbf{T}$ and $\mathbf{S}$ as well as $\mathbf{D}$ and $\mathbf{H}$, with a $(-1)$ phase for every $\mathbf{D}$ present. Note that, as we are at half filling, there is always the same number of $\mathbf{D}$ and $\mathbf{H}$, so $\hat{C}=\hat{C}^\dagger$ and $\hat{C}^2=\mathbf{1}$.
As it also anti-commutes with $\hat{Q}$, $\hat{C}$ exchanges the sectors with $Q=q$ and $Q=-q$. For $M$-even, we have a sector with $q=0$; in that sector the spectrum is symmetric around $E{=}0$ because of $\hat{C}$, however, this is not the case in other sectors.

{\bf Experimental protocols.---}In the main text, we used a quench protocol to observe the many-body scar on the superconducting quantum processor. This includes three main steps: state preparation, interaction and measurement. In the experimental sequence, we first initialize all the qubits, $Q_i$, in their ground state at their idle frequency $\omega_j$, the vertical couplers at around their sweet spots, and the horizontal couplers at around the frequency where the coupling strength between two nearest qubits are zero. The initial states are prepared by alternating the single-qubit gates and the two-qubit controlled-$\pi$ (CZ) gates. The preparation of the first family of scarred state $\ket{\Pi}$, which is a product state, is realized by applying single-qubit  XY rotations to every qubit. The preparation of the second family of scarred state $\ket{\phi_L}$ differs from the first family, as it is a ``cat'' state. We alternate single- and three two-qubit gates on the former four qubits, and apply XY rotations on the other qubits (as illustrated in Fig.~4 of the main text), with each coupler dynamically switched between nearly off and on to realize the CZ gate. $\mathrm{CNOT}$ gates in the experiment circuits are realized by $\mathrm{CZ}$ gate with Hadamard gate operated on the target qubit before and after.
 
 In the interaction step, we bias all qubits on resonance at the interaction frequency $\omega_i$. Meanwhile, the couplers are also tuned to turn on the net couplings between neighboring qubits.  After waiting for an interaction time $t$, we finally tune all qubit frequencies away from the interacting regime in order to measure the relevant quantities at their readout frequencies. Repeating this process with varying $t$ allows us to obtain the dynamics of the system. Also, using the quantum state tomography technique, we obtain the reduced density matrix of the subsystem $A$. 

\begin{figure}[bt]
\centering
\includegraphics[width=\linewidth]{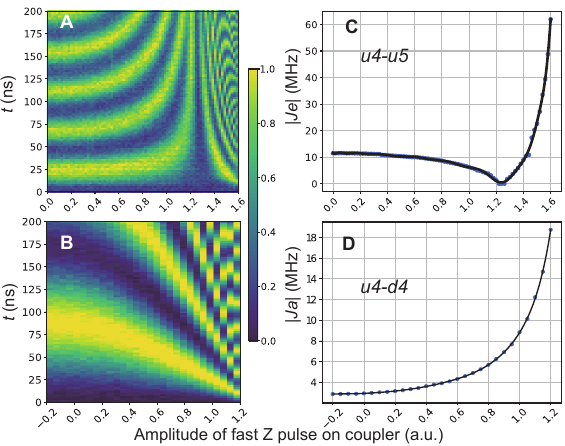}
\caption{{\bf Effective tunable couplings in the experimental device.} Swapping dynamics as tuned by the coupler for ({\bf A}) the two adjacent horizontal qubits ($u_4-u_5$) and ({\bf B}) the longitudinal qubits ($u_4-u_4$) is shown. The absolute effective coupling strengths as a function of the coupler Z pulse amplitude are obtained by fitting the oscillations. The horizontal coupling strength ({\bf C}) can be adjusted from positive to negative values, while the longitudinal coupling strength ({\bf D}) can only be tuned in the negative range.}
\label{fig:TunableCoupling}
\end{figure}

In experiment, tuning the coupling strengths of all couplers and tuning all qubits on resonance are two very important steps, for which we have conducted careful calibration following the procedure below:

1. Coarse-tune coupling strengths: the coupling strength between each pair of neighboring qubits is coarsely calibrated as a function of the amplitude of the coupler Z pulse, as shown in Fig.~\ref{fig:TunableCoupling}. In this process, we excite one of the qubits and then place them at $\omega_i$ for 200 ns, during which other qubits are placed 50 MHz above $\omega_i$, and other couplers are placed at around their maximum frequency. The coupling strength can be estimated by fitting the swapping dynamics. Following this process, we obtain the functions of all couplers.

2. Fine-tune coupling strengths: Keeping the frequencies of other qubits as above, we apply the corresponding Z pulse on other couplers according to the above results to roughly achieve the designed coupling strength. Then we slightly change the coupler Z pulse to fine-tune the coupling strength, validated by the swapping dynamics. This process is conducted for each pair of neighboring qubits.

3. Fine-tune frequencies of qubits: The frequencies of qubits may be affected by some factors such as Z pulse distortion, so we adopt the following strategies to minimize this uncertainty, during which all couplers are placed at their desired frequencies. We first apply $\pi$/2 pulse on one qubit, then place other qubits at a frequency above $\omega_i$ and tune this qubit to $\omega_i$ for a fixed delay, and finally measure its accumulated phase $\phi_{+}=\Delta{\omega_+}\times t$ by tomographic operations. We also measure its accumulated phase $\phi_{-}=\Delta{\omega_-}\times t$ when other qubits are placed 50 MHz below $\omega_i$. We then slightly change the qubit frequency to make $\phi_{+} + \phi_{-} \sim 0$. This process is implemented for every qubit.

{\bf Stability against perturbations.---}
The Hamiltonian describing our experimental device can be written as
 \begin{equation}\label{eq:expterms}
\begin{split}
    \hat{H}_\mathrm{exp}&= \hat{H}+\hat{H}_\mathrm{x} + \hat{V},\\
    \hat{H}_\mathrm{x} /2\pi  &=  J_{x}\sum_{k=1}^{M-1} \left( \hat{\mathbf{u}}^+_{k} \hat{\mathbf{d}}^-_{k+1} +\hat{\mathbf{d}}^+_{k} \hat{\mathbf{u}}^-_{k+1} + \mathrm{h.c.} \right), \\
    \hat{V}/2\pi  &= \frac{\eta}{2}  \sum_{k=1}^{M} \left(  \hat{\mathbf{u}}^+_k  \hat{\mathbf{u}}^+_k  \hat{\mathbf{u}}^-_k  \hat{\mathbf{u}}^-_k    + \hat{\mathbf{d}}^+_k  \hat{\mathbf{d}}^+_k  \hat{\mathbf{d}}^-_k  \hat{\mathbf{d}}^-_k   \right).
\end{split}
\end{equation}
Here, $\hat H$ denotes the Hamiltonian of Eq.~(1) in the main text, which has been reformulated in terms of bosons, where the standard raising and lowering operators are given by $\hat{\mathbf{u}}^\pm = \left( \hat{\mathbf{u}}^{x} \pm i\hat{\mathbf{u}}^{y} \right)/2$ (and, similarly, for $\hat{\mathbf{d}}^\pm$).  We consider a maximum of 2 photons per site.
The last two terms represent the  experimental imperfections due to the cross coupling $J_x$ between the diagonal qubits and nonlinearity $\eta$ of the transmon qubit~\cite{zhang2022many}. The precise values of the $J_x$ couplings as measured in our device are listed in the SM~\cite{SOM}. The terms in Eq.~(\ref{eq:expterms}) have been included in the numerical simulations presented in Figs.~\ref{fig:exp_scar1}-\ref{fig:exp_scar3} of the main text.

\begin{figure}[tb]
\centering
\includegraphics[width=\linewidth]{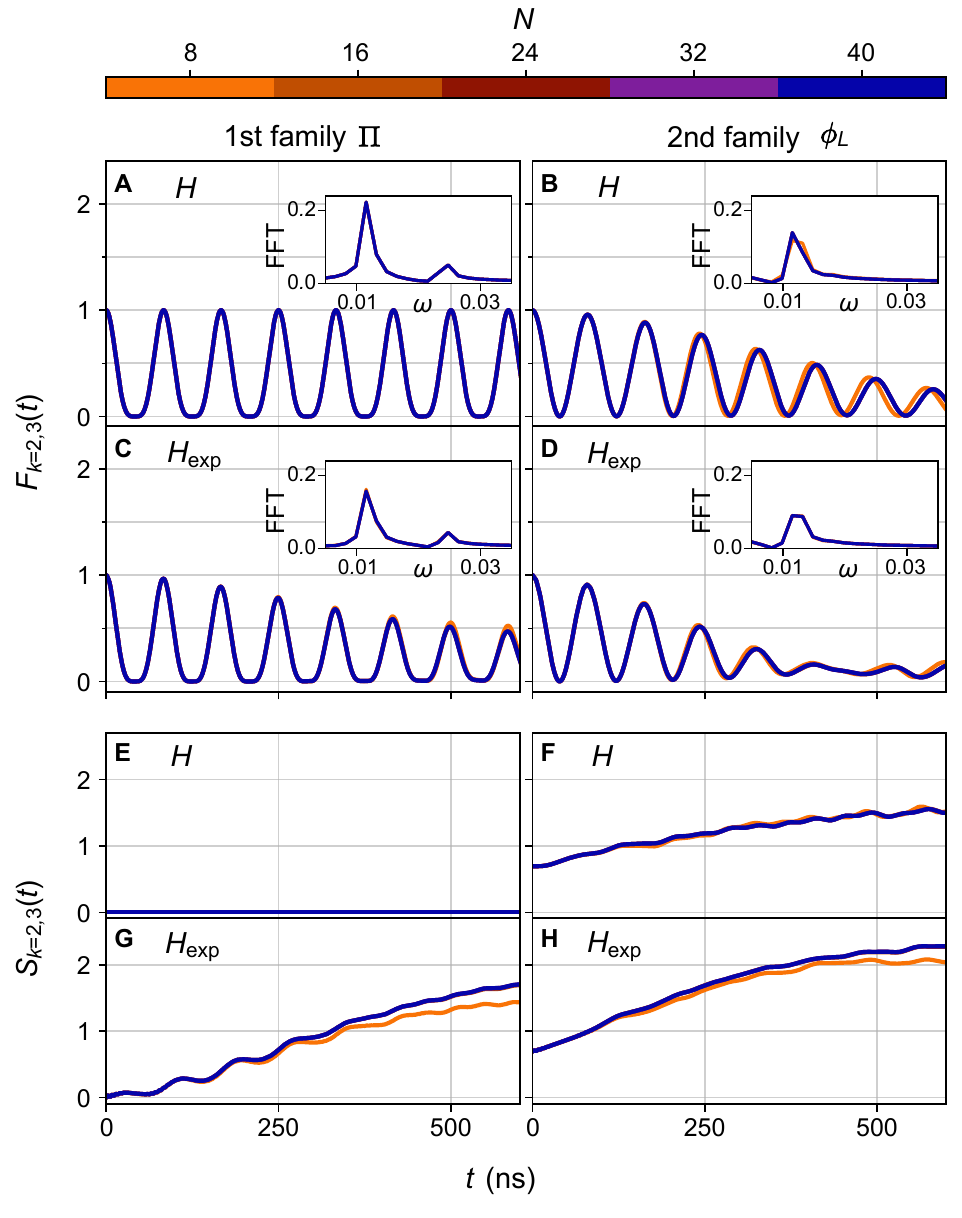}
\caption{{\bf Influence of experimental perturbations.} 
Fidelity and entanglement entropy dynamics of the subsystem $\{k=2,3\}$ obtained by the numerical simulations of the ideal Hamiltonian $\hat H$ in the main text (panels {\bf A},{\bf B},{\bf E} and {\bf F}) and the Hamiltonian~$\hat{H}_\mathrm{exp}$, which contains the perturbations in Eq.~(\ref{eq:expterms}) (panels {\bf C},{\bf D},{\bf G} and {\bf H}). Data for $N\geq 24$ is obtained using TDVP. The coupling parameters are identical to those in Fig.~4 in the main text and are $J_a=-6$ MHz, $J_{e,k}=-2$ MHz, $\eta=175$ MHz and $J_x \in [0.05,0.45]$ MHz.
}
\label{fig:exp_influence}
\end{figure}

The existence of $\hat H_x$ and $\hat  V$ terms weakens the amplitude of revival dynamics of both families of scar states. However, the perturbed model $\hat H_\mathrm{exp}$ still supports the two scar families and allows us to clearly distinguish between them. In Fig.~\ref{fig:exp_influence} we demonstrate the stability of our results with respect to these experimental perturbations in large systems $N \leq 40$ using matrix-product state methods and time-dependent variational principle (TDVP), as implemented in TenPy libraries~\cite{tenpy}. The accuracy of this calculation is controlled by the bond dimension of the matrix product states, and the convergence was ensured by requiring that the relative error in $F_{k=2,3}(t)$ and in $S_{k=2,3}(t)$ observables are always below $10^{-3}$ when comparing the two largest bond-dimensions used. Global quantities, such as many-body fidelity and bipartite entanglement entropy, were also monitored and showed good agreement between different bond dimensions. The strength of the cross couplings $J_x$ beyond the 16 first qubits have been randomly drawn from a uniform distribution in $[0.05,0.45]$ and are then kept identical across all system sizes. Consistent with the results in Fig.~4 in the main text, the fidelity dynamics of subsystem $A=\{k=2,3\}$ involves two frequencies for the first family of scar, one more than in the second family.  The initial entanglement entropy is $\ln{2}$ for the second family of scar, further distinguishing it from the first scar family which has zero initial value for the entropy. The data in Fig.~\ref{fig:exp_influence} reveals a remarkably fast convergence in system size. While the smallest system $N=8$ deviates slightly from larger sizes, we can see that  $F_{2,3}(t)$ and $S_{2,3}(t)$ are already fully converged across the experimentally accessible time interval of 500ns for system sizes $N\geq 16$.
\\

\bibliography{references}

\vspace*{0.4cm}

\noindent {\bf Acknowledgments}\\

\noindent We acknowledge Prof. Haohua Wang for supporting facilities that cover full experiments of this work from device fabrication to measurement. The device was fabricated at the Micro-Nano Fabrication Center of Zhejiang University. 
\\

\noindent {\bf Funding}\\

\noindent We acknowledge the support of the National Natural Science Foundation of China (Grant Nos. U20A2076,  12274367) and the National Key R\&D Program of China (Grant No. 2022YFA1404203). J.-Y.D. and Z.P.~acknowledge support by EPSRC grant EP/R513258/1 and by the Leverhulme Trust Research Leadership Award RL-2019-015. Statement of compliance with EPSRC policy framework on research data: This publication is theoretical work that does not require supporting research data. This research was supported in part by the National Science Foundation under Grant No. NSF PHY-1748958.\\

\noindent {\bf Author Contributions Statement}\\

\noindent L.Y., Z.P., and J.-Y.D formulated the theoretical proposal and performed the numerical simulations; H.D. and Y.G. performed the experiments and data analysis supervised by Z.W.; H.L. fabricated the device; Z.P., L.Y., H.D. and J.-Y.D. co-wrote the manuscript with inputs from other authors. All authors contributed to the experimental setup, discussions of the results and development of the manuscript.\\

\noindent {\bf Competing Interests Statement}\\

\noindent The authors declare no competing interests. \\

\noindent {\bf Data and Materials Availability}\\

\noindent 
All data needed to evaluate the conclusions in the paper are present in 
the paper and the Supplementary Materials. Supporting data are available at DOI:10.5281/zenodo.8433624.

\noindent 

\cleardoublepage

\beginsupplement

\onecolumngrid

\begin{center}
{
	\bf \large 
 Supplementary Material for ``Disorder-tunable entanglement at infinite temperature''
}

\vspace*{0.5cm}

Hang Dong$^{1}$, Jean-Yves Desaules$^{2}$, Yu Gao$^{1}$, Ning Wang$^{1}$, Zexian Guo$^{1}$, Jiachen Chen$^{1}$, Yiren Zou$^{1}$, Feitong Jin$^{1}$, Xuhao Zhu$^{1}$, Pengfei Zhang$^{1}$, Hekang Li$^{1}$, Zhen Wang$^{1}$, Qiujiang Guo$^{1}$, Junxiang Zhang$^{1}$, Lei Ying$^{1}$, Zlatko Papi\'c$^{2}$

\vspace*{0.2cm}
{\footnotesize \em
$^{1}$School of Physics, and ZJU-Hangzhou Global Scientific and Technological Innovation Center, \\
Zhejiang University, Hangzhou 310027, China\\
$^{2}$School of Physics and Astronomy, University of Leeds, Leeds, UK.
}

\end{center}

\vspace*{0.02cm}

\twocolumngrid

\setcounter{equation}{0}
\setcounter{figure}{0}
\setcounter{table}{0}
\setcounter{page}{1}
\setcounter{section}{0}

\section{Device information}

Our experiment is performed on a flip-chip superconducting quantum processor hosting $2\times20$ frequency-tunable transmon qubits in a ladder configuration. Each pair of adjacent qubits (both in rungs and side rails) are coupled by a tunable transmon coupler. All the control lines and readout resonators are located on a silicon substrate (bottom chip) and all the qubits and couplers are located on a sapphire substrate (top chip). The substrates are connected together with indium bumps as described in Ref.~\cite{zhang2022digital}. Each qubit can be controlled by three control lines. The microwave (XY) line can control the state of the qubit, while the flux (DC/Z) lines can change the frequency of the qubit. The maximum resonance frequencies for the qubits are around $4.6$ GHz and can be effectively tuned to below $4.0$ GHz. The qubit can be measured using a capacitively coupled readout resonator, with the frequency of the resonator ranging from $6.5$ GHz to $6.7$ GHz. Each coupler is equipped with its own flux (DC/Z) lines to tune its frequency, with a designed maximum frequency of around 9 GHz. The total coupling strength of each pair of adjacent qubits is composed of direct coupling between them and indirect coupling through the coupler. The former is dominated by the direct capacitance between the qubits while the latter is determined by the capacitance between the qubit and the coupler. Both have been carefully designed so as to realize different tunable ranges in the rungs and side rails. Detailed information about the processor can be found in Tab.~\ref{tab:dev_para_hamiltonian}, including the idle frequencies, average single-qubit gate error, energy relaxation times, and dephasing times. Note that in the experiment, the qubits form a coupled system that is insensitive to each qubit's flux noise. Thus effective dephasing times are usually much longer than the Ramsey dephasing time $T^*_{2}$. For example, using the spin echo technique, the dephasing time is typically measured to be around $10~\mu$s. Finally, the estimated values of  the $J_x$ cross couplings are provided in Tab.~\ref{tab:Jx}.

\begin{table} [htb]
    \centering
    \addtolength{\tabcolsep}{+2pt}
\caption{\small {\bf Parameters of the processor.} $\omega^{0}_j$ is the maximum frequency of $Q_j$, known as the sweet spot; $\omega^{\text{i}}_j$ is the idle frequency of $Q_j$ where single-qubit XY rotations are applied and the average single-qubit gate Pauli error $e_{sq}$ is measured via simultaneous cross entropy benchmarking (XEB). All qubits are biased to $\omega^{\text{I}} \approx 4.375$~GHz to activate the interaction, where the energy relaxation time $T_{1, j}$ and the Ramsey dephasing time $T^*_{2, j}$ of each qubit $Q_j$ are measured. 
}
\begin{tabular}{l|c|cc|ccc}
        \hline
        \hline
         Qubit & $\omega^{0}_j$ (GHz) & $\omega^{\text{i}}_j$ (GHz) & $e_{sq}$ ($\%$) & $T_{1, j}$ ($\mu$s) & $T^*_{2, j}$ ($\mu$s) \\
        \hline
        $u_{1  }$&   4.650&     4.357&    0.532&      46.3&     1.29\\
        $d_{1  }$&   4.578&     3.982&    0.442&      24.0&     1.37\\
        $u_{2  }$&   4.650&     4.074&    0.619&      20.6&     1.39\\
        $d_{2  }$&   4.670&     4.304&    0.620&      26.5&     0.77\\
        $u_{3  }$&   4.684&     4.344&    0.718&      25.5&     1.09\\
        $d_{3  }$&   4.639&     3.996&    0.647&      30.7&     1.21\\
        $u_{4  }$&   4.745&     4.068&    0.973&      32.2&     0.97\\
        $d_{4  }$&   4.619&     4.309&    0.589&      17.4&     1.58\\
        $u_{5  }$&   4.684&     4.347&    1.456&      25.8&     0.89\\
        $d_{5 }$&   4.622&     4.029&    0.760&      25.3&     1.30\\
        $u_{6 }$&   4.596&     4.044&    0.635&     33.1&     2.00\\
        $d_{6 }$&   4.460&     4.334&    0.475&      23.6&     1.05\\
        $u_{7 }$&   4.541&     4.314&    0.976&      26.6&     1.72\\
        $d_{7 }$&      -     &     4.064&    1.412&      15.9&     0.84\\
        $u_{8 }$&      -     &     4.022&    0.684&      13.6&     2.35\\
        $d_{8 }$&   4.569&     4.349&    0.451&      38.6&     1.59\\
        \hline
        Ave. &       -&         -&    0.749&      26.6&     1.34\\
        \hline
\end{tabular}
\label{tab:dev_para_hamiltonian}
\end{table}

\begin{table}[bt]
    \centering
    \addtolength{\tabcolsep}{+2pt}
\caption{\small {\bf Cross coupling strengths.} List of the cross couplings $J_x$ from experimental measurements.
}
\begin{tabular}{c|c||c|c}
        \hline
        \hline
         Pairs & $J_x$ (MHz) & Pairs  & $J_x$ (MHz) \\
        \hline
        $u_{1}-d_2$&   0.19&     $d_{1}-u_2$&    0.10\\
        $u_{2}-d_3$&   0.30&     $d_{2}-u_3$&    0.34\\
        $u_{3}-d_4$&   0.37&     $d_{3}-u_4$&    0.45\\
        $u_{5}-d_5$&   0.31&     $d_{4}-u_5$&    0.33\\
        $u_{6}-d_6$&   0.25&     $d_{5}-u_6$&    0.26\\
        $u_{7}-d_7$&   0.19&     $d_{6}-u_7$&    0.18\\
        $u_{8}-d_8$&   0.28&     $d_{7}-u_8$&    0.16\\
        \hline
\end{tabular}
\label{tab:Jx}
\end{table}

\section{Finite-size scaling}
\begin{figure}[bt]
\centering
\includegraphics[width=\linewidth]{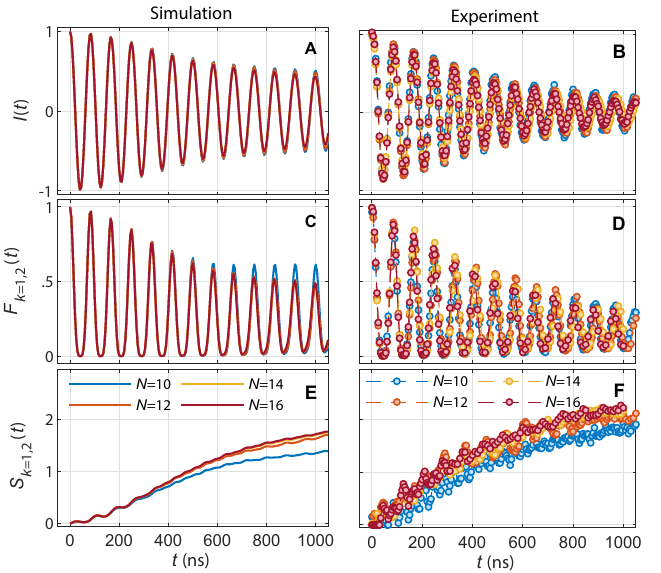}
\caption{{\bf System size scaling of dynamics.} Dynamics of imbalance ({\bf A}-{\bf B}), subsystem fidelity ({\bf C}-{\bf D}), and entanglement entropy ({\bf E}-{\bf F}) for simulations and experiments, respectively, with different system sizes ranging from $N=10$ to $16$. Data is for the first type of scar, with the same parameters as those used in Fig.~3 of the main text.
}
\label{fig:scaling}
\end{figure}

In Fig.~\ref{fig:scaling} we show the experimental results of imbalance, subsystem fidelity, and entanglement entropy of the first four qubits in system sizes ranging from $N = 10$ to $16$ qubits obtained using exact diagonalization. The experimental results are in good agreement with the numerical simulations, complementing similar analysis in the Methods section.

\section{Subsystem fidelity for two families of scars}

In general, the fidelity of a subsystem $A$ for both pure and mixed initial states is given by
 \begin{equation}
     F_A(t) = \left(  \mathrm{tr} \sqrt{   \sqrt{\rho_A(t)}   \rho_A(0) 
 \sqrt{\rho_A(t)}  }   \right)^2.
 \end{equation}
The reduced density matrix is given by
\begin{equation}
\begin{split}
     \rho_A(t) &= \mathrm{tr}_B \rho(t) =  \mathrm{tr}_B  \Big( \sum_{n,n^\prime} c_n(t)c^\ast_{n^\prime}(t)  | n\rangle   \langle n^\prime|   \Big)   \\
     &= \sum_{\beta,n,n^\prime}    \Big( c_n(t)c^\ast_{n^\prime}(t)  \langle b_\beta | b_n  \rangle   \langle b_{n^\prime} | b_\beta  \rangle   \Big)   | a_{n}\rangle  \langle a_{n^\prime}|,
\end{split}
\end{equation}
where $A=\{a_1,\cdots,a_\alpha,\cdots\}$ and $B=\{b_1,\cdots,b_\beta,\cdots\}$ represent the subsystems $A$ and $B$, respectively. Here, $\alpha,\beta=1,\cdots,D_{A,B}$ and $n,n^\prime=1,\cdots,D$ denote the indices of subsystems and the entire system, respectively, and $D,D_{A,B}$ are the dimensions of entire Hilbert space and subspace $A$ and $B$.

\begin{figure}[tb]
\centering
\includegraphics[width=\linewidth]{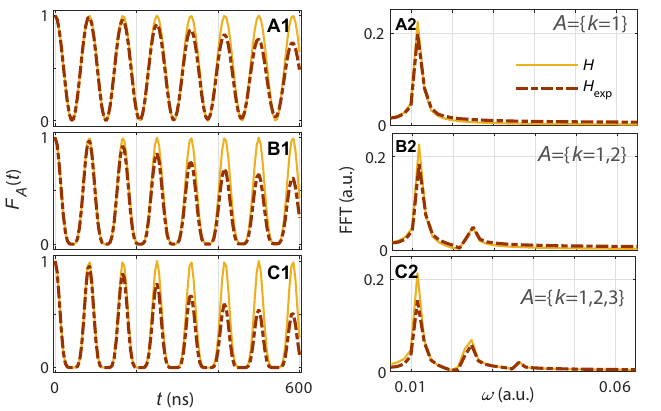}
\caption{{\bf Scaling of fidelity with subsystem size.} Subsystem fidelity dynamics of the initial state $|\Pi\rangle$ for Hamiltonians $\hat{H}$ in the  main text and $\hat{H}_\mathrm{exp}$ in Eq.~(\ref{eq:expterms}) for subsystem sizes of $2,4,6$, respectively. System size is $N=10$ and the coupling parameters are $J_a=J_e$, $J_x=0.1J_a$.}
\label{fig:subsystem1}
\end{figure}

The first scar family can be probed by studying the dynamics of $|\Pi\rangle$, which is constrained to a hypercube subspace with a single particle in each dimer $\{u_k,d_k\}$. Then, the dynamics of the reduced density matrix is simplified to 
\begin{equation}
\begin{split}
    \rho_A(\Pi;t) &= \sum_{\beta,n,n^\prime \in \mathrm{S}}    \Big( c_n(t)c^\ast_{n^\prime}(t)  \langle b_\beta | b_n  \rangle   \langle b_{n^\prime} | b_\beta  \rangle   \Big)   | a_{n}\rangle  \langle a_{n^\prime}|  \\
    &=   2^{N_B/2}  \sum_{\alpha,\alpha^\prime \in \mathrm{S_A}}    c_\alpha(t) c^\ast_{\alpha^\prime}(t) | a_{\alpha}\rangle  \langle a_{\alpha^\prime}|    .
\end{split}
\end{equation}
Thus, the reduced density matrix of the subsystem $A$ for $|\Pi\rangle$ is equivalent to an isolated system of size $N_A$. For the subsystem $A=\{k=1\}$, its fidelity is a cosine function with only one peak in its Fourier spectrum. As the subsystem size increases, the revival fidelity consists of multiple frequencies in general. This result is confirmed by numerical simulations, as shown in Fig~\ref{fig:subsystem1}. 

 \begin{figure}[tb]
\centering
\includegraphics[width=\linewidth]{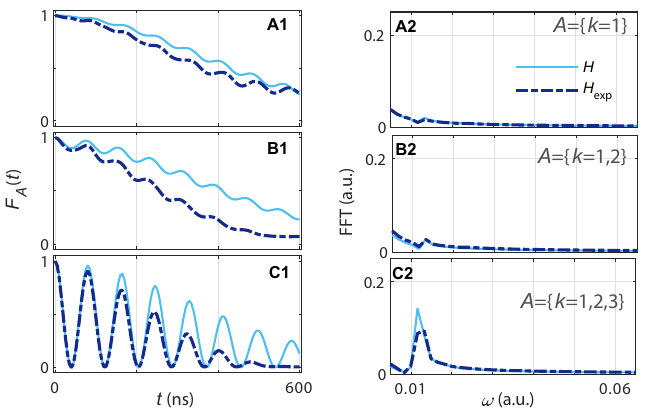}
\caption{{\bf Fidelity dynamics.} Subsystem fidelity dynamics of the initial state $|\phi_L\rangle$ for the Hamiltonian $\hat{H}$ in the main text and $\hat{H}_\mathrm{exp}$ in Eq.~(\ref{eq:expterms}) for subsystem sizes of $2,4,6$. The parameters are identical to those in Fig.~\ref{fig:subsystem1}.}
\label{fig:subsystem2}
\end{figure}

For the initial state $|\phi_L\rangle$, the fidelity revivals are not exact even for the ideal Hamiltonian $\hat H$. In this case, the reduced matrix cannot be simplified as $\rho_A(\Pi;t)$ above. However, to understand the fidelity dynamics it is helpful to consider a small system. For $A=\{k=1\}$ and $\{k=1,2\}$, the corresponding initial state are $|\phi_L\rangle=\left(\ket{{\bullet \atop \bullet}}+\ket{{\circ \atop \circ}}\right)/\sqrt{2}$ and $\left(\ket{{\bullet\circ \atop \bullet\circ}}+\ket{{\circ\bullet \atop \circ\bullet}}\right)/\sqrt{2}$, which are entangled states. Their fidelity dynamics have no revivals. If the subsystem $A$ involves the third dimers, i.e. $A=\{k=1,2,3\}$, the initial state $|\phi_L\rangle=\left(1/\sqrt{2}\right)\left(\ket{{\bullet\circ \atop \bullet\circ}}+\ket{{\circ\bullet \atop \circ\bullet}}\right)\otimes \ket{{\bullet \atop \circ}}$ contains a $\ket{{\bullet \atop \circ}}$. Its fidelity dynamics have one-frequency revival dynamics, as shown in Fig.~\ref{fig:subsystem2}. 

Furthermore, we emphasize that the experimental perturbations do not affect our distinction between states $\Pi$ and $\phi_L$. In Figs.~\ref{fig:subsystem1} and~\ref{fig:subsystem2}, we numerically compute the fidelity dynamics for different subsystem sizes based on the experimental Hamiltonian. The signatures that distinguish the first and second families of scars can be clearly observed. 

\section{Hamiltonian in the dimer basis}\label{app:H_dim}

The Hamiltonian from the main text can be re-written as 
\begin{equation}
\begin{split}
    \hat{H} = \sum_{k=1}^{M-1} \hat{h}_{k,k+1}^\|+\sum_{k=1}^{M} \hat{h}_{k}^\perp   \label{eq:H2}
\end{split}
\end{equation}
with
\begin{equation}
\begin{split}
    \hat{h}_{k,k+1}^\| =&   \frac{J_{e,k}}{2}\left(\hat{\mathbf{u}}^x_k\hat{\mathbf{u}}^x_{k+1}{+}\hat{\mathbf{u}}^y_k\hat{\mathbf{u}}^y_{k+1}\hspace{-0.1cm}{-}\hat{\mathbf{d}}^x_k\hat{\mathbf{d}}^x_{k+1}\hspace{-0.1cm}{-}\hat{\mathbf{d}}^y_k\hat{\mathbf{d}}^y_{k+1}\right), \quad \\
     \hat h^\perp_k =& \frac{J_a}{2} \left(\hat{\mathbf{u}}^x_k\hat{\mathbf{d}}^x_k+\hat{\mathbf{u}}^y_k\hat{\mathbf{d}}^y_k \right)+\omega_k \left(\hat{\mathbf{u}}^z_k+\hat{\mathbf{d}}^z_k\right).
\end{split}
\end{equation}
The $\hat{\mathbf{u}}^\alpha$ and $\hat{\mathbf{d}}^\alpha$ are Pauli matrices acting on the top and bottom rows, respectively. The state $\bullet$ denotes an up-spin and $\circ$ denotes a down-spin, such that $\hat{\mathbf{u}}\ket{\circ}=-\ket{\circ}$ and $\hat{\mathbf{u}}\ket{\bullet}=\ket{\bullet}$.
The effect of applying the Hamiltonian on any state in the computational ($Z$) basis is straightforward. However, its effect on the dimer basis is not obvious and we explicitly work it out in this section. To achieve this, it will be convenient to use the above decomposition of the Hamiltonian as a sum of terms that either act on a single dimer or on two dimers at once. The former corresponds to diagonal terms and hopping perpendicular to the ladder (denoted by $\hat h^\perp_k$), while the latter encompasses hopping parallel to the ladder and is denoted by $\hat h_{k,k+1}^\|$.  

In the main text, we introduced the convenient local dimer basis, spanned by doublon $\ket{\mathbf{D}}$, holon $\ket{\mathbf{H}}$, triplet $\ket{\mathbf{T}}$ and singlet $\ket{\mathbf{S}}$ states. Those were given by:
\begin{equation}\label{eq:dimerbasis}
\begin{split}
 \ket{\mathbf{D}}&:=\ket{{\bullet \atop \bullet}}, \quad \ket{\mathbf{H}}:=\ket{{\circ \atop \circ}}, \\
 \ket{\mathbf{T}}&:=\frac{1}{\sqrt{2}}\left(\ket{{\circ \atop \bullet}}+\ket{{\bullet \atop \circ}}\right),  \\ 
\ket{\mathbf{S}}&:=\frac{1}{\sqrt{2}}\left(\ket{{\circ \atop \bullet}}-\ket{{\bullet \atop \circ}}\right).    
\end{split}
\end{equation}
 It is straightforward to see that every state in the dimer basis is an eigenstate of $\hat h^{\perp}_k$ with
\begin{equation}
    \begin{aligned}
    \hat h^{\perp}_k\ket{\mathbf{H}}&=-2\omega_k\ket{\mathbf{H}}, \\
    \hat h^{\perp}_k\ket{\mathbf{T}}&=\ \  J_a\ket{\mathbf{T}}, \\
    \hat h^{\perp}_k\ket{\mathbf{D}}&= 2\omega_k\ket{\mathbf{D}}, \\
    \hat h^{\perp}_k\ket{\mathbf{S}}&= -J_a\ket{\mathbf{S}}.
    \end{aligned}
\end{equation}
The action of $\hat{h}_{k,k+1}^{||}$ is more complicated and requires looking at two neighboring dimers. It is straightforward to show $\hat h^{\|}_{k,k+1}\ket{\sigma\sigma}=0$ for any $\sigma=\mathbf{H},\mathbf{D},\mathbf{T},\mathbf{S}$. The non-zero terms are
\begin{equation}
    \begin{aligned}
    \hat h^{\|}_{k,k+1}\ket{\mathbf{TS}}&=J_e(\ket{\mathbf{HD}}-\ket{\mathbf{DH}}), \\
    \hat h^{\|}_{k,k+1}\ket{\mathbf{TH}}&=\; -J_e\ket{\mathbf{HS}}, \\
    \hat h^{\|}_{k,k+1}\ket{\mathbf{SH}}&=\; -J_e\ket{\mathbf{HT}}, \\
    \hat h^{\|}_{k,k+1}\ket{\mathbf{TD}}&=J_e\ket{\mathbf{DS}}, \\
    \hat h^{\|}_{k,k+1}\ket{\mathbf{SD}}&=J_e\ket{\mathbf{DT}}, \\
    \hat h^{\|}_{k,k+1}\ket{\mathbf{HD}}&=J_e(\ket{\mathbf{TS}}-\ket{\mathbf{ST}}). \\
    \end{aligned}
\end{equation}
The missing combinations can be obtained by flipping the two sites under consideration on the right- and left-hand sides. Importantly, this leads to the following combinations of states being annihilated by $\hat h^{\|}_{k,k+1}$:
\begin{equation}
    \begin{aligned}
    \hat h^{\|}_{k,k+1}\left(\ket{\mathbf{TS}}+\ket{\mathbf{ST}}\right)&=0, \\
    \hat h^{\|}_{k,k+1}\left(\ket{\mathbf{HD}}+\ket{\mathbf{DH}}\right)&=0.
    \end{aligned}
\end{equation}

\section{Structure of scarred eigenstates}

The rainbow scar construction~\cite{Langlett2022,Wildeboer22} is based on two subsystems (labeled ``1'' and ``2''), with respective Hamiltonians obeying the relation
\begin{equation}\label{eq:rainbowcondition}
\hat{H}_2=-\mathcal{M}\hat{H}_1^\star\mathcal{M}^\dagger,
\end{equation}
As the spectra of two subsystems are identical up to a minus sign, the composite system has a large zero-energy subspace, spanned by pairs of eigenstates with energies $E$ and $-E$. Provided that $\mathcal{M}$ maps each site in subsystem 1 to a single site in subsystem 2, as in Eq.~(\ref{eq:rainbowcondition}) (i.e., with no global term mixing all sites), then the zero-energy subspace contains a rainbow state, 
\begin{equation}
\ket{\mathbf{I}}{=}\frac{1}{\sqrt{\mathcal{D}_1}}\bigotimes_{j=1}^{M}\sum_{\sigma_j}\ket{\sigma_j}{\otimes} ({\mathcal{M}}\ket{\sigma_j}),
\end{equation}
where $\mathcal{D}_1$ denotes the number of states in the subsystem 1 and $\sigma_j$ denotes the state of site $j$ in it. The state $\ket{\mathbf{I}}$ is simply a tensor product of Bell pairs, and it has maximal entanglement between the two subsystems. Importantly, $\ket{\mathbf{I}}$ is an eigenstate of the full system, \emph{independently} of the microscopic details of $\hat{H}_1$~\cite{Langlett2022}. 
One  can then make the two subsystems interact by adding a  term $\hat{H}_\mathrm{int}$. If it is chosen such that it has $\ket{\mathbf{I}}$ as an eigenstate, $\ket{\mathbf{I}}$ will be a scarred eigenstate of the full system while other states generically become ergodic.

In our case, subsystem 1 is the top row of the ladder while subsystem 2 is the bottom row. The transformation related to their Hamiltonians is 
\begin{equation}
\mathcal{M}=\left(\prod_{k=1}^M  \hat{\mathbf{d}}_k^x\right)\hat{\mathcal{P}}_{\mathbf{d}\leftrightarrow \mathbf{u}},
\end{equation}
where $\hat{\mathcal{P}}_{\mathbf{d}\leftrightarrow \mathbf{u}}$ is the operator exchanging the top and bottom row and the term between parentheses is a particle-hole exchange. The rainbow state can then be written succinctly as
\begin{equation}
\ket{\mathbf{I}}{\equiv}\ket{E_M}{=}\bigotimes_{j=1}^{M}\frac{1}{\sqrt{2}}\left(\ket{\bullet \atop \circ}+\ket{\circ \atop \bullet}\right) \equiv \ket{\mathbf{TTT}\ldots \mathbf{TT}},
\end{equation}
which reveals the structure of Bell pairs formed between the two rungs of the ladder. Upon addition of $\hat{H}_\mathrm{int}$, the rainbow state $\ket{I}$ remains an exact eigenstate with energy $MJ_a$, thus becoming a rainbow scar of the full model. 

The previous construction is not limited to a single state. We can generate additional rainbow scar states by acting on $\ket{\mathbf{I}}$ with an operator $\hat{O}\otimes \mathbb{1}_2$, where $\hat{O}$ commutes with the Hamiltonian $\hat{H}_1$~\cite{Langlett2022}.  The resulting state, by construction, still belongs to the zero-energy subspace of the combination of $\hat{H}_1$ and $\hat{H}_2$. Provided that the resulting state is also an eigenstate of $\hat{H}_\mathrm{int}$, it is then a scarred eigenstate of the full system. Two different families of scars in our model in Eq.~(\ref{eq:model}) can be built this way by choosing different operators $\hat{O}$. Moreover, it allows to get simple product states with overlap only on these special eigenstates.

\subsection{First scar family}

To construct the first scar family, we use the operator $\hat{Z}=\sum_{k=1}^M \hat{\mathbf{u}}^z_k$. As explained in the main text, this operator commutes with $\hat H_1$ and it can be applied up to $M$ times, changing triplets into singlets and vice versa. The resulting states, $\ket{E_n}$, are members of the first type of scar family. They represent a symmetric superposition of all configurations with a fixed number of singlets and triplets, and it is straightforward to verify that they are eigenstates of the model with energy $E_n=J_a (2n-M)$. 

The underlying structure of the first family of scarred states $\ket{E_n}$ is the restricted spectrum generating algebra~\cite{MotrunichTowers}. This is seen by noting that they can be built 
using the raising operator $\hat S^+=\sum_{k=1}^M \ket{T_k}\bra{S_k}$. Thus, we can equivalently express the first type of scar states as
\begin{eqnarray}
   \ket{E_n} = \frac{1}{\mathcal{N}}  (\hat S^+)^n \ket{E_0} = \frac{1}{\mathcal{N}}(\hat S^+)^n\ket{\mathbf{SS}\ldots \mathbf{S}}, 
\end{eqnarray}
where $\mathcal{N}$ is a normalization constant. We recognize that these scarred eigenstates are built in a similar way to previously known examples, e.g., in the spin-1 XY model~\cite{Iadecola2019_2}. In Sec.~\ref{app:entanglement} we compute their entanglement entropy for the bipartition perpendicular to the ladder, and we find that the eigenstate at zero energy has $S_{1,\perp}=0.5+0.5\log(\pi M/8)$, in the limit $M\to \infty$. We note that the su(2) algebraic structure of these states implies they have extensive multipartite entanglement~\cite{DesaulesQFI,Dooley2023}.

\subsection{Second scar family}
The second scar family can be built recursively from the first family, where the role of the operator $\hat O$ in the rainbow construction is played by the subsystem Hamiltonian $\hat H_1$. Specifically, the second type of scar states are given by raising from  $\ket{E_{n-1}}$ according to 
\begin{equation}
    \ket{E^\prime_n}{\propto}\hat{P}^Q_{M{-}2n}\left[\hat{H}_1{-}\left(\sum_k \frac{\omega_k}{M}\right)\hat{Z}\right]\ket{E_{n-1}},
\end{equation}
or, equivalently, by lowering from $\ket{E_{n+1}}$ as 
\begin{equation}
    \ket{E^\prime_n}{\propto}\hat{P}^Q_{M{-}2n}\left[\hat{H}_1{-}\left(\sum_k \frac{\omega_k}{M}\right)\hat{Z}\right]\ket{E_{n+1}}.
\end{equation}
In these expressions, $\hat P_q^{Q}$ is a projector on the sector of $\hat Q$ with
eigenvalue $q$.  

In order to write the states $\ket{E_n^\prime}$ in a more explicit form, let us define ensembles of sites $\Lambda=\{1,2,\ldots M\}$, $\Lambda_k=\{k,k+1\}$ and $\Lambda_{\overline{k}}=\Lambda \setminus \Lambda_k$. 
With this, we can express $\ket{E^\prime_n}$ as
\begin{equation}\label{eq:2ndscarwf}
\begin{aligned}
\ket{E^\prime_n}\hspace{-0.1cm}&=\hspace{-0.15cm}\sum_{k=1}^{M-1}\hspace{-0.15cm}\frac{J_{e,k}}{2\mathcal{N}_n}\left(\ket{\mathbf{HD}}_{\Lambda_k}\hspace{-0.09cm}{+}\ket{\mathbf{DH}}_{\Lambda_k}\right)\hspace{-0.1cm}\otimes \hspace{-0.1cm}\ket{M{-}n{-}1,n{-}1}_{\overline{\Lambda}_k} \\
&+\frac{1}{\mathcal{N}_n}\left(\sum_{k=1}^{M}\hat{T}_k\omg_k\right) \ket{M-n,n}_{\Lambda},
\end{aligned}
\end{equation}
with $n=1, 2, \ldots, M-1$, $\omg_j=\omega_j-\frac{1}{M}\sum_{k=1}^M \omega_k$,  and $\ket{a,b}_A$ the symmetric superposition of all states with $a$ singlets and $b$ triplets on sites in the ensemble $A$. More details about the construction of these states and the proof that they are eigenstates of the model are given in Secs.~\ref{app:building}-\ref{app:proof}, where we also derive the normalization factors $\mathcal{N}_n$. There, we prove that $\ket{E_n^\prime}$ have the same energies as the first family of scar states, i.e., $E_n^\prime = J_a (2n-M)$.

To the best of our knowledge, there is no su(2) algebraic construction for the second family of scarred states  $\ket{E^\prime_n}$, despite their equal energy spacing. Indeed, the latter is generated by acting with an operator on $\ket{E_n}$, and not on $\ket{E^\prime_{n-1}}$ or $\ket{E^\prime_{n+1}}$. 
Moreover, the second type of scarred eigenstates depend sensitively on the parameters $J_{e,k}$ and $\omega_k$ of the model. Therefore, the entanglement of these states is not fixed, like in the first family of scars, as we discuss in detail in Sec.~\ref{app:entanglement}. For a cut perpendicular to the ladder, the second type of scarred states generally possess entanglement entropy that scales logarithmically with subsystem size. While we are unable to rule out the possibility of volume-law entanglement scaling, our analytical and numerical results strongly suggest that, in the large $M$ limit, the second type of scarred state with zero energy obeys $S_{\perp,1}<S_{\perp,2}<S_{\perp,1}+\log 4$~\cite{SOM}. This suggests it should always be possible to find a low-entangled state that overlap only on the scarred stats of the second family, regardless of the values of parameters. By tuning the hopping $J_{e,k}$ one can then change which state has this anomalous overlap, as we will demonstrate in Sec.~\ref{app:dynamical}. 

\subsection{Conditions on the Hamiltonian parameters}
With some insights into the structure of both families of scarred eigenstates, we can now check if some conditions on the Hamiltonian parameters could potentially be relaxed. As the only parameters with no disorder are the $J_a$, it is natural to ask if we are allowed to introduce inhomogeneity in these couplings while preserving the constructed scarred states. 
Randomizing $J_a$ can be easily shown not to work. Recall from the rainbow scar construction, the scarred state $\ket{E_n}$ of the first family is simply the symmetric superposition of all combinations with $M-n$ singlets and $n$ triplets. In order for these states to be eigenstates of the full system, they also need to be eigenstates of $\hat{H}_\mathrm{int}$, which encompasses the action of the vertical XY terms. The states $\ket{SSSS\ldots}$ and $\ket{TTTT\ldots}$ are eigenstates even if the $J_a$ are disordered, but their energy will change. However, the other eigenstates will generally \emph{not} be eigenstates of $\hat{H}_\mathrm{int}$ unless the $J_a$ are equal. 

Indeed, let us take the state with 3 dimers, $\frac{1}{2}\left(\ket{STT}{+}\ket{TST}{+}\ket{TTS}\right)$, as an example. Applying $\hat{H}_\mathrm{int}$ to it leads to a different prefactor for each term in the superposition. The first term will have a prefactor of $-J_{a,1}+J_{a,2}+J_{a,3}$, the second term a prefactor of $J_{a,1}-J_{a,2}+J_{a,3}$, and so on. This state is therefore not an eigenstate of $\hat{H}_\mathrm{int}$ unless all the $J_a$ are equal. Importantly, it is also not possible to create eigenstates of $\hat{H}_\mathrm{int}$ by linear combinations of the $\ket{E_n}$. 
The same reasoning holds for the second family of scarred states. In the end, if the $J_a$ are disordered, the model only has two exact scarred eigenstates, $\ket{E_0}$ and $\ket{E_N}$, which are incidentally the only states in the sectors $Q=-N/2$ and $Q=N/2$. 

Other ways of relaxing the constraints on the Hamiltonian parameters are less obvious as the $J_{e,k}$ and $\omega_k$ are already disordered. Since the rainbow scars construction rests on the similarity between the top and bottom chains, it is not possible to use different disorder realizations for each. However, one might ask if perhaps a different mirror transformation $\hat{\mathcal{M}}$ relating the two rows could lead to a simpler Hamiltonian. In particular, as it is difficult to engineer a superconducting chip capable of both strong positive and negative couplings, it would be desirable to have positive $J_{e}$ coupling on both chains. In that case, these two parts of the Hamiltonian can satisfy the rainbow scar construction by using
\begin{equation}\label{eq:newmirror}
\mathcal{M}=\left( \hat{\mathbf{d}}_1^x\hat{\mathbf{d}}_2^y\hat{\mathbf{d}}_3^x\hat{\mathbf{d}}_4^y\hat{\mathbf{d}}_5^x\ldots \hat{\mathbf{d}}_M^y\right)\hat{\mathcal{P}}_{\mathbf{d}\leftrightarrow \mathbf{u}},
\end{equation}
and the system governed by $\hat{H}_\mathbf{u}+\hat{H}_\mathbf{d}$ will also have rainbow scars in that case. However, the key difference is the structure of the scarred eigenstates and what happens when they are acted upon by the vertical XY terms of $\hat{H}_\mathrm{int}$. 

Due to the nature of the mirror transformation in Eq.~(\ref{eq:newmirror}), the rainbow state will now be equal to $\ket{TSTS\ldots}$, up to an irrelevant overall phase. We can then create the first family of scars by acting on this state with the total magnetization on the top chain $\hat{Z}$ that exchanges triplets and singlets, as these states will also be zero-energy eigenstates of $\hat{H}_\mathbf{u}+\hat{H}_\mathbf{d}$. Starting from the state $\ket{TSTS\ldots}$ and generating a basis of this subspace using a Lanczos procedure, we find the following states in a simple example with $M=3$:
\begin{equation}
    \begin{aligned}
        \ket{S_0}&=\ket{TST} \\
        \ket{S_1}&=\frac{1}{\sqrt{3}}\left(\ket{SST}+\ket{TTT}+\ket{TSS} \right)\\
        \ket{S_2}&=\frac{1}{\sqrt{3}}\left(\ket{TTS}+\ket{SSS}+\ket{STT}\right)\\
        \ket{S_3}&=\ket{STS}.
    \end{aligned}
\end{equation}
These states will be eigenstates of the full system only if they are eigenstates of $\hat{H}_\mathrm{int}$. A quick check shows that this is only the case for $\ket{S_0}$ and $\ket{S_3}$. For all other states, the terms in the superposition have a different number of triplets and singlets, leading to a different prefactor. As the states are also non-overlapping, it is not possible to create eigenstates of $\hat{H}_\mathrm{int}$ by linear combinations of the $\ket{S_n}$. Thus, in the full system, only $\ket{S_0}$ and $\ket{S_M}$ survive  as exact eigenstates. In fact, these exact states were previously found in Ref.~\cite{IadecolaLocalization} which studied the model with lower and upper $J_e$ equal. 
One way to reinstate the $\ket{S_n}$ as eigenstates would be by staggering the $J_a$ between odd and even bonds. However, this does not solve the problem of having large positive and negative couplings. The resulting model would also be equivalent to the one we study, up to a rotation around the Z axis on every other spin of the lower chain.

In summary, there are crucial conditions on the couplings on the model that are due to the important role of $\hat{H}_\mathrm{int}$ that connects the two copies of the system. While the prerequisite of the rainbow scar construction -- an opposite spectrum between two halves of a system -- is met in many systems, the additional condition that these eigenstates be preserved by $\hat{H}_\mathrm{int}$ is usually what precludes the existence of this type of scars. Additionally, acting on these states with $\hat{H}_\mathbf{u}$ will always lead to other zero-energy eigenstates of $\hat{H}_\mathbf{u}+\hat{H}_\mathbf{d}$, but it is even harder to ensure that such states would be eigenstates of $\hat{H}_\mathrm{int}$ (especially in the presence of disorder). To the best of our knowledge, our model is the first where this has been shown to occur.

\section{Building scarred states of the second family}\label{app:building}

In this section, we derive the exact form of the scarred states of the second family. The formal proof that these states are eigenstates of the model is delegated to the subsequent section. 

In order to reveal the structure of the second family of scarred states, let us first derive the action of $\hat{H}_1{-}\left(\sum_k \omega_k/M \right)\hat{Z}$ in the dimer basis. First, we define
\begin{equation}
\hat h^{\|,1}_{k,k+1}=\frac{J_{e,k}}{2}\left(\hat{\mathbf{u}}^x_k\hat{\mathbf{u}}^x_{k+1}{+}\hat{\mathbf{u}}^y_k\hat{\mathbf{u}}^y_{k+1}\right).
\end{equation}
From there it is straightforward to see that 
\begin{equation}
    \hat{H}_1{-}\left(\sum_k \frac{\omega_k}{M}\right)\hat{Z}=\sum_{k=1}^{M-1} \hat{h}_{k,k+1}^{\|,1}+\sum_{k=1}^{M} \omg_k \mathbf{u}_k^z,
\end{equation}
where $\omg_k=\omega_k{-}\left(\sum_i \omega_i/M\right)$.
As we will only apply this operator to scarred states of the first family, which contain no $\ket{\mathbf{D}}$ or $\ket{\mathbf{H}}$, we can ignore any configurations containing them.
The action of $\hat h^{\|,1}_{k,k+1}$ and $\mathbf{u}^z_k$ on dimers is then
\begin{equation}
    \begin{aligned}
    \hat h^{\|,1}_{k,k+1}\ket{\mathbf{TT}}&=\frac{1}{2}J_e(\ket{\mathbf{HD}}+\ket{\mathbf{DH}}), \\
    \hat h^{\|,1}_{k,k+1}\ket{\mathbf{SS}}&=-\frac{1}{2}J_e(\ket{\mathbf{HD}}+\ket{\mathbf{DH}}), \\
    \hat h^{\|,1}_{k,k+1}\ket{\mathbf{\mathbf{TS}}}&=\frac{1}{2}J_e(\ket{\mathbf{HD}}-\ket{\mathbf{DH}}), \\
    \hat h^{\|,1}_{k,k+1}\ket{\mathbf{ST}}&=-\frac{1}{2}J_e(\ket{\mathbf{HD}}-\ket{\mathbf{DH}}), \\
    \omg_k\hat{\mathbf{u}}^z_k\ket{\mathbf{S}}&=-\omg_k\ket{\mathbf{T}}, \\
    \omg_k\hat{\mathbf{u}}^z_k\ket{\mathbf{T}}&=-\omg_k\ket{\mathbf{S}}.
    \end{aligned}
\end{equation}
From there, we immediately see that 
\begin{equation}\label{eq:cancellation}
 \hat h^{\|,1}_{k,k+1}(\ket{\mathbf{TS}}+\ket{\mathbf{ST}})=0.   
\end{equation}

To represent the symmetric superposition, we recall the notation introduced to describe scarred states of the first family:
\begin{equation}
    \ket{E_n}=\frac{1}{\mathcal{N}}\ket{L{-}n,n}=\frac{1}{\mathcal{N}}\sum_{\ket{\phi}  \in (L{-}n,n) }\ket{\phi}.
\end{equation}
This state is a fully symmetric superposition of all configurations on $L$ sites with $L-n$ singlets and $n$ triplets. For brevity, we will first work out what happens when we apply  $\hat{h}^{\|,1}_{1,2}$ (the same is true for any $\hat{h}^{\|,1}_{k,k+1}$):
\begin{equation}
\begin{aligned}
\hat{h}^{\|,1}_{1,2}\ket{E_{n-1}}=&\frac{\hat{h}^{\|,1}_{1,2}}{\mathcal{N}}\ket{M{-}n{+}1,n{-}1}\\
=&\frac{\hat{h}^{\|,1}_{1,2}}{\mathcal{N}}\Big[(\ket{\mathbf{TS}}+\ket{\mathbf{ST}})\ket{M{-}n,n{-}2}\\
&+\ket{\mathbf{TT}}\ket{M{-}n{+}1,n{-}3}\\
&+\ket{\mathbf{SS}}\ket{M{-}n{-}1,n{-}1}\Big] \\
=&\frac{J_{e,1}}{2\mathcal{N}}(\ket{\mathbf{HD}}{+}\ket{\mathbf{DH}})\ket{M{-}n{+}1,n{-}3} \\
&{-}\frac{J_{e,1}}{2\mathcal{N}}(\ket{\mathbf{HD}}{+}\ket{\mathbf{DH}})\ket{M{-}n{-}1,n{-}1},
\end{aligned}
\end{equation}
where we used the condition (\ref{eq:cancellation}) to cancel the contribution of the first term. Applying the projector $\hat{P}^Q_{M{-}2n}$ singles out one of the terms:
\begin{equation}\label{eq:hpar1_rai}
\begin{aligned}
\hat{P}^Q_{M{-}2n}\hat{h}^{\|,1}_{1,2}\ket{E_{n-1}}={-}\frac{J_{e,1}}{2\mathcal{N}}(\ket{\mathbf{HD}}{+}\ket{\mathbf{DH}})\ket{M{-}n{-}1,n{-}1}.
\end{aligned}
\end{equation}
Next, we look at the action of $\omg_1\hat{\mathbf{u}}^z_1$:
\begin{equation}
\begin{aligned}
\omg_1\hat{\mathbf{u}}^z_1\ket{E_{n-1}}&{=}\frac{\omg_1\hat{\mathbf{u}}^z_1}{\mathcal{N}}\ket{M{-}n{+}1,n{-}1}\\
&{=}\frac{\omg_1\hat{\mathbf{u}}^z_1}{\mathcal{N}}\Big[\ket{\mathbf{T}}\ket{M{-}n{+}1,n{-}2}{+}\ket{\mathbf{S}}\ket{M{-}n,n{-}1}\Big]\\
&{=}\frac{-\omg_1}{\mathcal{N}}\Big[\ket{\mathbf{S}}\ket{M{-}n{+}1,n{-}2}{+}\ket{\mathbf{T}}\ket{M{-}n,n{-}1}\Big].
\end{aligned}
\end{equation}
Applying the projector in this case gives 
\begin{equation}
\hat{P}^Q_{M{-}2n}\omg_1\hat{\mathbf{u}}^z_1\ket{E_{n-1}}{=}-\frac{\omg_1}{\mathcal{N}}\ket{\mathbf{T}}\ket{M{-}n,n{-}1}.
\end{equation}
The result is similar if we act on another site $k$, with a prefactor $\omg_k$ and a triplet on that site. Ultimately, we end up with a collection of all states with $n$ singlets and $n-1$ triplets, but each of them has a prefactor that depends on the location of the triplets. Let us introduce the operator $\hat{T}_k$ that gives 1 if this site is a triplet and 0 otherwise. We can then write 
\begin{equation}\label{eq:omg1_rai}
\hat{P}^Q_{M{-}2n}\sum_{k=1}^M\omg_k\hat{\mathbf{u}}^z_k\ket{E_{n-1}}{=}-\frac{1}{\mathcal{N}}\left(\sum_{k=1}^{M}\hat{T}_k\omg_k\right) \ket{M{-}n,n}.
\end{equation}

To write down the scarred states of the second family, we now simply need to gather the terms from Eqs.~(\ref{eq:hpar1_rai}) and (\ref{eq:omg1_rai}). We also remove the overall minus sign and normalize the state. Let us introduce the ensembles of sites $\Lambda=\{1,2,\ldots M\}$ , $\Lambda_k=\{k,k+1\}$ and $\Lambda_{\overline{k}}=\Lambda-\Lambda_k=\{1,2\ldots k-1,k+2,k+3,\ldots M\}$. They represent, respectively, all sites, sites $k$ and $k+1$, and all sites except $k$ and $k+1$. This allows us to write $\ket{E^\prime_n}$ as
\begin{equation}\label{eq:scar2_wf}
\begin{aligned}
\ket{E^\prime_n}\hspace{-0.1cm}&= \hspace{-0.15cm}\sum_{k=1}^{M-1}\hspace{-0.15cm}\frac{J_{e,k}}{2\mathcal{N}_n}(\ket{\mathbf{HD}}_{\Lambda_k}\hspace{-0.09cm}{+}\ket{\mathbf{DH}}_{\Lambda_k})\hspace{-0.1cm}\otimes \hspace{-0.1cm}\ket{M{-}n{-}1,n{-}1}_{\overline{\Lambda}_k} \\
&+\frac{1}{\mathcal{N}_n}\left(\sum_{k=1}^{M}\hat{T}_k\omg_k\right) \ket{M-n,n}_{\Lambda},
\end{aligned}
\end{equation}
with $n=1, 2, \ldots, M-1$. As an example, let us write out the case for $M=4$ and $n=1$:
\begin{equation}
\begin{aligned}
\ket{E_1^\prime}=\frac{1}{2\mathcal{N}_1}\Big[&J_{e,1}(\ket{\mathbf{DHSS}}+\ket{\mathbf{HDSS}})+J_{e,2}(\ket{\mathbf{SDHS}}\\
&+\ket{\mathbf{SHDS}})+J_{e,3}(\ket{\mathbf{SSDH}} + \ket{\mathbf{SSHD}})\Big]\\
+\frac{1}{\mathcal{N}_1}\Big[&\omg_1\ket{\mathbf{TSSS}}+\omg_2\ket{\mathbf{STSS}}\\
&+\omg_3\ket{\mathbf{SSTS}}+\omg_4\ket{\mathbf{SSST}} \Big].
\end{aligned}
\end{equation}

Finally, let us show that the same result is obtained if we generate $\ket{E^\prime_n}$ by lowering from $\ket{E_{n+1}}$. We have
\begin{equation}\label{eq:hpar1_low}
\begin{aligned}
\hat{P}^Q_{M{-}2n}\hat{h}^{\|,1}_{1,2}\ket{E_{n+1}}=\frac{J_{e,1}}{2\mathcal{N}}(\ket{\mathbf{HD}}{+}\ket{\mathbf{DH}})\ket{M{-}n{-}1,n{-}1}.
\end{aligned}
\end{equation}
and 
\begin{equation}
\hat{P}^Q_{M{-}2n}\omg_1\hat{\mathbf{u}}^z_1\ket{E_{n+1}}{=}\frac{-\omg_1}{\mathcal{N}}\ket{\mathbf{S}}\ket{M{-}n{-}1,n}.
\end{equation}
From the latter equation, we can derive that 
\begin{equation}
\hat{P}^Q_{M{-}2n}\sum_{k=1}^M\omg_k\hat{\mathbf{u}}^z_k\ket{E_{n-1}}{=}\frac{-1}{\mathcal{N}}\left(\sum_{k=1}^{M}\hat{S}_k\omg_k\right) \ket{M{-}n{-}1,n},
\end{equation}
where $\hat{S}_k$ gives 1 if this site is a triplet and 0 otherwise. Now we can notice that $\ket{M{-}n{-}1,n}$ is composed entirely of triplets and singlets. Consequently,  $(\hat{T}_k+\hat{S}_k)=\mathbf{1}_k$ when acting on that state. Moreover,  we know that the $\omg$ must sum to 0 by construction. This allows us to state that
\begin{equation}
     \sum_{k=1}^M\omg_k (\hat{S}_k{+}\hat{T}_k)\ket{M{-}n{-}1,n}=\hspace{-0.13cm}\sum_{k=1}^M\omg_k\ket{M{-}n{-}1,n}{=}0.
\end{equation}
From there we can conclude that 
\begin{equation}\label{eq:omg1_low}
\begin{aligned}
\hat{P}^Q_{M{-}2n}\sum_{k=1}^M\omg_k\hat{\mathbf{u}}^z_k\ket{E_{n-1}}&{=}\frac{-1}{\mathcal{N}}\left(\sum_{k=1}^{M}\hat{S}_k\omg_k\right) \ket{M{-}n{-}1,n} \\
&=\frac{1}{\mathcal{N}}\left(\sum_{k=1}^{M}\hat{T}_k\omg_k\right) \ket{M{-}n{-}1,n}.
\end{aligned}
\end{equation}
Gathering the results of Eqs.~(\ref{eq:hpar1_low}) and (\ref{eq:omg1_low}) we find the same result as in Eq.~(\ref{eq:scar2_wf}).

\subsection{Normalization}
As we have the exact wavefunction for the scarred states of the second family, we can compute their normalization factor $\mathcal{N}_n$. It admits a simple expression 
\begin{equation}\label{eq:Nn}
    \mathcal{N}_n=\sqrt{\binom{M-2}{n-1}\left[\frac{1}{2}\sum_{k=1}^{M-1}J_{e,k}^2+\sum_{k=1}^{M}\omg_k^2\right]}.
\end{equation}
This expression does not contain any cross-term $\omg_k \omg_j$ because all possible combinations of $k\neq j$ appear and we can then express them as
\begin{equation}
\sum_{k=1}^{M-1}\sum_{j=k+1}^M \omg_k\omg_j=-\frac{1}{2}\sum_{k=1}^M \omg_k^2,
\end{equation}
by using the fact that $\sum_{k=1}^M \omg_k=0$ and as such
\begin{equation}
0=\left(\sum_{k=1}^M \omg_k\right)^2=\sum_{k=1}^M \omg_k^2+2\sum_{k=1}^{M-1}\sum_{j=k+1}^M \omg_k\omg_j.
\end{equation}

\section{Proof that scarred states are eigenstates}\label{app:proof}

In this section, we prove that the two families of scarred states, written down in the main text, are eigenstates of the model in Eq.~(\ref{eq:H2}). We first address the straightforward cases of the first family of scars with $n=0$ and $n=M$. We then show the proof for the slightly more complicated $n=1$ case and finally demonstrate that the same arguments generalize to arbitrary $n$.

\subsection{\texorpdfstring{$n=0$ and $n=M$}{n=0 and n=M} scarred states}

Proving that $\ket{E_0}=\ket{SS\ldots S}$ is an eigenstate is trivial, as we know that $\ket{\mathbf{SS}}$ is an eigenstate of $\hat{h}^{\|}_{k,k+1}$ with energy 0 and $\ket{\mathbf{S}}$ is an eigenstates of $\hat{h}^{\perp}_{k}$ with energy $-J_a$. Thus, $\ket{E_0}$ must be an eigenstate of $\hat{H}$ with energy $-MJ_a$. Similarly, $\ket{\mathbf{TT}}$ is an eigenstate of $\hat{h}^{\|}_{k,k+1}$ with energy 0 and $\ket{\mathbf{T}}$ is an eigenstate of $\hat{h}^{\perp}_{k}$ with energy $J_a$. Thus, $\ket{E_M}=\ket{\mathbf{TT}\ldots \mathbf{T}}$ must be an eigenstate of $\hat{H}$ with energy $M J_a$.

\subsection{\texorpdfstring{$n=1$}{n=1} scarred states}\label{app:E1_prime}

For $n=1$ we will prove the eigenstate property by considering a $2\times 3$ ladder and then show that the same holds for larger systems. Consider the state 
\begin{equation}
\begin{aligned}
\ket{\psi}&=\beta_1 \left(\ket{\mathbf{HDS}}+\ket{\mathbf{DHS}}\right)+\beta_2 \left(\ket{\mathbf{SHD}}+\ket{\mathbf{SDH}}\right)\\
&+\alpha_1 \ket{\mathbf{TSS}}+\alpha_2\ket{\mathbf{STS}}+\alpha_3\ket{\mathbf{SST}}.
\end{aligned}
\end{equation}
Applying the Hamiltonian to this state and after some algebra, we obtain
\begin{equation}
\begin{aligned}
\hat H\ket{\psi}
&=-J_a\ket{\psi} \\
&+\left[ 2\beta_1\left( \omega_1{-}\omega_2\right) {+}J_{e,1}\left(  \alpha_2{-}\alpha_1\right) \right] \left( \ket{\mathbf{DHS}}{-}\ket{\mathbf{HDS}}\right) \\
&+\left[ 2\beta_2\left( \omega_2{-}\omega_3\right) {+}J_{e,2}\left(  \alpha_3{-}\alpha_2\right) \right] \left( \ket{\mathbf{SDH}}{-}\ket{\mathbf{SHD}}\right) \\
&+\left( J_{e,2}\beta_1-J_{e,1}\beta_2 \right)\left(\ket{\mathbf{HTD}}-\ket{\mathbf{DTH}} \right).
\end{aligned}
\end{equation}
For $\ket{\psi}$ to be an eigenstate, the coefficients must obey
\begin{align}
    2\beta_1 \left( \omega_1{-}\omega_2\right) {+}J_{e,1}\left(  \alpha_2{-}\alpha_1\right)=0, \label{eq:eqsys1} \\
    2\beta_2 \left( \omega_2{-}\omega_3\right) {+}J_{e,2}\left(  \alpha_3{-}\alpha_2\right)=0, \label{eq:eqsys2} \\
    J_{e,2}\beta_1-J_{e,1}\beta_2=0.\label{eq:eqsys3}
\end{align}
In general, we have 5 unknowns but only 3 equations, leaving room for two non-trivial solutions. 
The first option is to set all $\alpha_j$ to be equal and all $\beta_j$ to zero:
\begin{equation}\label{eq:ab1}
    \alpha_j=1, \quad \beta_j=0.
\end{equation}
This corresponds to the scarred state of the first family, $\ket{E_1}$. The other solution is given by the scarred states of the second family that obey (up to a normalization factor)
\begin{equation}\label{eq:ab2}
    \alpha_j=\omega_j-\frac{1}{M}\sum_k \omega_k=\omg_j, \quad   \beta_j=\frac{J_{e,j}}{2}.
\end{equation}
Furthermore, we can verify that the two families of scars are orthogonal as their overlap is given by 
\begin{equation}
    \sum_{j} \omg_j=\sum_j\left(\omega_j-\frac{1}{M}\sum_k \omega_k\right)=0.
\end{equation}
This completes the proof for the special case of $M=3$. However, generalizing this to an arbitrary $M$ is now straightforward. The general state is  
\begin{equation}
\begin{aligned}
    \ket{\psi}&=\beta_1 \left(\ket{\mathbf{HDS}\ldots \mathbf{S}}+\ket{\mathbf{DHS}\ldots \mathbf{S}}\right)\\
    &+\beta_2 \left(\ket{\mathbf{S}\ldots \mathbf{SHDS}}+\ket{\mathbf{S}\ldots \mathbf{SDHS}}\right)\\
    &+\ldots+\beta_{M-2} \left(\ket{\mathbf{S}\ldots \mathbf{SHD}}+\ket{\mathbf{S}\ldots \mathbf{SDH}}\right)\\
&+\beta_{M-1} \left(\ket{\mathbf{HDS}\ldots \mathbf{S}}+\ket{\mathbf{DHS}\ldots \mathbf{S}}\right)\\
&+\alpha_1 \ket{\mathbf{TS}\ldots \mathbf{S}}+\alpha_2 \ket{\mathbf{STS}\ldots \mathbf{S}}+\ldots\\
&+\alpha_{M-1} \ket{\mathbf{S}\ldots \mathbf{STS}}+\alpha_M \ket{\mathbf{S}\ldots \mathbf{ST}}.
\end{aligned}
\end{equation}
Now we have $M$ different $\alpha_j$ and $M-1$ different $\beta_j$, so $2M-1$ unknowns in total. For each block of $2\times 2$ sites we get an equation similar to Eqs.~\eqref{eq:eqsys1} and \eqref{eq:eqsys2}. Thus, for $j=1$ to $M-1$ we have
\begin{equation}\label{eq:eq_alpha_LS}
    2\beta_j \left( \omega_{j}{-}\omega_{j+1}\right) {+}J_{e,j}\left(  \alpha_{j+1}{-}\alpha_j\right)=0.
\end{equation}
For each rectangular block of $2\times 3$ sites we get an equation similar to Eq.~\eqref{eq:eqsys3}. Hence, for $j=1$ to $M-2$ we have
\begin{equation}\label{eq:eq_beta_LS}
    J_{e,j+1}\beta_j-J_{e,j}\beta_j+1=0.
\end{equation}
For a chain with $2M$ sites, this yields $(M-1)+(M-2)=2M-3$ equations. So for any system size we always get at least two scarred states with energy $-(M-2)J_a$.
It is easy to check that the two solutions given in Eqs.~\eqref{eq:ab1}
and \eqref{eq:ab2} are still valid.

\subsection{Other values \texorpdfstring{$n$}{n}}
While we treated the case $n=1$ on its own to provide a simple example, the recipe is exactly the same for general $n$ (except for $n=0$ and $n=M$ that were already proven).
We use the same Ansatz in which we only consider states with one $DH+HD$ pair in a background of $n-1$ triplets and $M-n-1$ singlets and states with $n$ triplets and $M-n$ singlets. 

We first restrict our investigation to an arbitrary location $k,\ k+1$ for the hole-doublon pair. 
\begin{equation}
\begin{aligned}
    \ket{\psi}=\beta_j\left(\ket{\ldots \mathbf{XHDY}\ldots}+\ket{\ldots \mathbf{XDHY}\ldots}\right)\\
    +\alpha_{TS}\ket{\ldots \mathbf{XTSY}\ldots} + \alpha_{ST}\ket{\ldots \mathbf{XSTY}\ldots},
\end{aligned}
\end{equation}
where $\mathbf{X}$ and $\mathbf{Y}$ denote either $\mathbf{S}$ or $\mathbf{T}$.
For $\hat{h}^\perp$, the contribution on all other sites except $k$ and $k+1$ is diagonal and equal to $J_a(2n-M)$. Therefore, to prove that states are eigenstates with this energy, we need to prove that the action of the rest of the Hamiltonian annihilates the state.  
For $\hat{h}^{\|}_{k,k+1}$ we only have to care about the action on sites $k-1$ to $k+2$. Indeed, the rest of the state is composed of triplets and singlets. For any other pair, if it is $\mathbf{SS}$ or $\mathbf{TT}$, then it is annihilated by the action of the Hamiltonian. If, instead, it is $\mathbf{TS}$, then there exists another state in the superposition, with the same weight, that has $\mathbf{ST}$ instead. Therefore, their superposition is also annihilated by the Hamiltonian. 

First, we can look at what happens if we act on the middle pair. This leads to 
\begin{equation}
\begin{aligned}
    2\beta_j\left(\omega_j-\omega_{j+1} \right)\left(\ket{\ldots \mathbf{XDHY}\ldots}-\ket{\ldots \mathbf{XHDY}\ldots}\right)\\
    +J_{e,j}\left(\alpha_{ST}-\alpha_{TS}\right)\left(\ket{\ldots \mathbf{XDHY}\ldots}-\ket{\ldots \mathbf{XHDY}\ldots}\right),
\end{aligned}
\end{equation}
and so
\begin{equation}\label{eq:alpha_n}
2\beta_j\left(\omega_{j}-\omega_{j+1} \right)    +J_{e,j}\left(\alpha_{ST}-\alpha_{TS}\right)=0.
\end{equation}
We get a unique equation for every of the $M-1$ pair of sites and for every of the $\binom{M-2}{n-1}$ possible background configurations, where $n=1$ to $M-1$ is the index of the scarred states. 

Now we still need to look at the effect of $\hat{h}^{\|}$ on $\mathbf{XD}$, $\mathbf{XH}$ (as well as $\mathbf{HY}$ and $\mathbf{DY}$). For that, we need to also consider the $\mathbf{DH}$ and $\mathbf{HD}$ pair placed one site to the left:
\begin{equation}
\begin{aligned}
    \beta_{j-1}\left(\ket{\ldots \mathbf{HDXY}\ldots}+\ket{\ldots \mathbf{DHXY}\ldots}\right)\\
    +\beta_j\left(\ket{\ldots \mathbf{XHDY}\ldots}+\ket{\ldots \mathbf{XDHY}\ldots}\right).
    \end{aligned}
\end{equation}
Applying the Hamiltonian to these states leads to 
\begin{equation}
\begin{aligned}
    \beta_{j-1}J_{e,j}\left(\ket{\ldots H\overline{\mathbf{X}}\mathbf{DY}\ldots}-\ket{\ldots D\overline{\mathbf{X}}\mathbf{HY}\ldots}\right)\\
    -\beta_j J_{e,j-1}\left(\ket{\ldots H\overline{\mathbf{X}}\mathbf{DY}\ldots}+\ket{\ldots D\overline{\mathbf{X}}\mathbf{HY}\ldots}\right),
\end{aligned}
\end{equation}
where $\overline{\mathbf{X}}=\mathbf{T}$ if $\mathbf{X}=\mathbf{S}$ and  $\overline{\mathbf{X}}=\mathbf{S}$ is $\mathbf{X}=\mathbf{T}$. Hence we get an equation
\begin{equation}\label{eq:eq_beta}
    \beta_{j-1}J_{e,j}-\beta_j J_{e,j-1}=0,
\end{equation}
for $j=1$ to $M-2$. Thus, for the $\beta$ coefficients, we always have $M-1$ unknown and $M-2$ equations.  Furthermore, these equations take the form of Eq.~\eqref{eq:eq_beta} and are \emph{identical} for any value of $n$.
Once again we recognize that setting all $\alpha$ equal and all $\beta$ to 0 is a valid solution, and so the first family of scarred state is indeed an eigenstate. 
As for the $n=1$, we also recognize that Eq.~\eqref{eq:eq_beta} admits $\beta_j=J_{e,j}/2$ as a solution. 

For the $\alpha$, if we add a contribution of $\omg_j$ for each site that has a triplet on site $j$ as in the second family of scars, we recognize that $\alpha_{ST}$ and $\alpha_{TS}$ have the same contributions outside of site $k$ and $k+1$. It is then easy to see that $\alpha_{ST}-\alpha_{TS}=\omg_{j+1}-\omg_{j}$ and that Eq.~\eqref{eq:alpha_n} is satisfied, concluding our proof.

\section{Dynamical signatures of two scar families}\label{app:dynamical}

Both families of scarred states are evenly spaced in energy with spacing $2J_a$, hence they can detected by persistent revivals following the quench from a suitably chosen initial state. In this section, we identify such initial states for the two scar families and prove the existence of revivals by computing the overlap of initial states with scarred eigenstates derived in previous sections. We illustrate how these results can be used to enhance quantum revivals by modulating the coupling between the qubits. 

\subsection{Reviving initial state for the first scar family}\label{app:revivals1}

To probe the first scar family, in the main text we used the state $\ket{\Pi}$ given by:
\begin{eqnarray}
    \ket{\Pi}=\ket{{\bullet \atop \circ}{\bullet \atop \circ}\cdots {\bullet \atop \circ}}.
\end{eqnarray}
This state must undergo perfect revivals due to the exact su(2) algebraic structure, as we now explain. From the raising operator $\hat{S}^+ = \sum_k \ket{T_k} \bra{S_k}$, we can infer $\hat{S}^x\propto \hat{S}^+ +(\hat{S}^+)^\dagger$ or, equivalently,
\begin{eqnarray}\label{eq:Sx}
 \hat{S}^x &\propto& 
\sum_k \left( \ket{{\circ \atop \bullet}} \bra{{\circ \atop \bullet}} -\ket{{\bullet \atop \circ}}\bra{{\bullet \atop \circ}} \right)_k.
\end{eqnarray}
The corresponding $\hat{S}^z$ operator is given by $\hat{S}^z = \sum_k \ket{T_k}\bra{T_k} - \ket{S_k}\bra{S_k}$, which is easily seen to have the same action as $\hat H$ in this subspace. Hence, if we prepare the system in the lowest-weight state of $\hat{S}^x$, it will undergo perfect precession around an effective field in the $z$-direction. It is easy to see that our state  $\ket{\Pi}$ is indeed the lowest weight state of Eq.~(\ref{eq:Sx}).
The precession that starts out in $\ket{\Pi}$ state will result in perfect state transfer to the highest-weight state $\ket{\Pi^\prime}=\ket{{\circ \atop \bullet}{\circ \atop \bullet}\cdots {\circ \atop \bullet}}$. 

The existence of revivals from the  $\ket{\Pi}$ initial state can be shown more explicitly by computing the overlap $|\langle \Pi | E_n \rangle|^2$ with eigenstates of the first scar family. First, it will be important to observe that we obtain a simple product state when all possible combinations of singlets and triplets are summed up:
\begin{equation}
    \sum_{n=0}^M \ket{M-n,n}=\bigotimes_{k=1}^M(\ket{\mathbf{T}}+\ket{\mathbf{S}})=2^{M/2}\ket{{\circ \atop \bullet}\cdots {\circ \atop \bullet}},
\end{equation}
where in the second equality we made use of
$\ket{\mathbf{T}}+\ket{\mathbf{S}}=\sqrt{2}\ket{{\circ \atop \bullet}}$. 
Since $\ket{E_n}=\binom{M}{n}^{-1/2} \ket{M-n,n}$, it is easy to see that
\begin{equation}
    \ket{\Pi^\prime}=\ket{{\circ \atop \bullet}\cdots {\circ \atop \bullet}}=\sum_{n=0}^M\sqrt{\frac{\binom{M}{n}}{2^M}}\ket{E_n}.
\end{equation}
The same procedure can applied to the $\ket{\Pi}$ state by noting that $\ket{\mathbf{T}}-\ket{\mathbf{S}}=\sqrt{2}\ket{{\bullet \atop \circ}}$. This means that there is now a factor of $-1$ for each singlet present, such that
\begin{equation}
    \ket{\Pi}=\ket{{\bullet \atop \circ}\cdots{\bullet \atop \circ}}=\sum_{n=0}^M(-1)^{M-n}\sqrt{\frac{\binom{M}{n}}{2^M}}\ket{E_n}.
\end{equation}
Thus, the $\ket{\Pi}$ and $\ket{\Pi'}$ states only have overlap on the first family of scarred eigenstates. As the latter are regularly spaced in energy, the dynamics initialized in $\ket{\Pi}$ or $\ket{\Pi'}$ exhibits perfect revivals.

\subsection{Reviving initial state for the second scar family}\label{app:revivals2}

For scarred states of the second family, the identification of the reviving initial state $\ket{\phi_L}$ is more subtle as the eigenstates now depend on disorder realization and there is no apparent algebraic structure. 
However, we can once again appeal to the fact that the sum of all singlet and triplet configurations can be written as a simple state in the Fock basis. For any $n$, $\ket{E_n^\prime}$ contain terms
\begin{equation}
\frac{J_{e,k}}{2\mathcal{N}_n}(\ket{\mathbf{HD}}_{\Lambda_k}\hspace{-0.09cm}{+}\ket{\mathbf{DH}}_{\Lambda_k})\hspace{-0.1cm}\otimes\hspace{-0.1cm}\ket{M{-}n{-}1,n{-}1}_{\Lambda_{\overline{k}}},
\end{equation}
with $\mathcal{N}_n$ the normalization factor given in Eq.~(\ref{eq:Nn}). Recall that $\Lambda_k$ denotes sites $k$ and $k+1$ and $\Lambda_{\overline{k}}$ denotes all sites except $k$ and $k+1$.
As a consequence, summing all $\ket{E^\prime_n}$ with a prefactor $\mathcal{N}_n$ gives
\begin{equation}
\begin{aligned}
&\sum_{n=1}^{M-1}\mathcal{N}_n\ket{E_n^\prime}\\
&=\sum_{k=1}^M\frac{J_{e,k}}{2}(\ket{\mathbf{HD}}_{\Lambda_k}\hspace{-0.09cm}{+}\ket{\mathbf{DH}}_{\Lambda_k})\hspace{-0.1cm}\otimes\hspace{-0.1cm}\sum_{n=1}^{m-1}\ket{M{-}n{-}1,n{-}1}_{\overline{\Lambda}_k} \\
&+\left(\sum_{k=1}^M \omg_k \hat{T}_k\right)\sum_{n=1}^{M-1}\ket{M-n,n}_{\Lambda} \\
&=\sum_{k=1}^M\frac{J_{e,k}}{2}(\ket{\mathbf{HD}}_{\Lambda_k}\hspace{-0.09cm}{+}\ket{\mathbf{DH}}_{\Lambda_k})\hspace{-0.1cm}\otimes2^{\frac{M{-}2}{2}}\hspace{-0.1cm}\ket{{\circ \atop \bullet}\cdots {\circ \atop \bullet}}_{\overline{\Lambda}_k} \\
&+\left(\sum_{k=1}^M \omg_k \hat{T}_k\right)\left(2^{M/2}\ket{\Pi^\prime}-\ket{T\cdots T}- \ket{S\cdots SS} \right) \\
&=2^{\frac{M{-}4}{2}}\sum_{k=1}^M J_{e,k}(\ket{\mathbf{HD}}_{\Lambda_k}\hspace{-0.09cm}{+}\ket{\mathbf{DH}}_{\Lambda_k})\hspace{-0.1cm}\otimes\hspace{-0.1cm}\ket{{\circ \atop \bullet}\cdots {\circ \atop \bullet}}_{\overline{\Lambda}_k} \\
&+2^{M/2}\left(\sum_{k=1}^M \omg_k \hat{T}_k\right)\ket{\Pi^\prime}.
\end{aligned}
\end{equation}
In order to simplify this state and to make the orthogonality with $\ket{E_n}$ obvious, we will consider the initial state 
\begin{equation}
\begin{aligned}
    \ket{\phi^\prime_J}=&\frac{1}{\mathcal{Z}}\sum_{k=1}^M J_{e,k}(\ket{\mathbf{HD}}_{\Lambda_k}\hspace{-0.09cm}{+}\ket{\mathbf{DH}}_{\Lambda_k})\hspace{-0.1cm}\otimes\hspace{-0.1cm}\ket{{\bullet \atop \circ}\cdots {\bullet \atop \circ}}_{\overline{\Lambda}_k} \\
    =&\frac{1}{\mathcal{Z}}\Bigg( J_{e,1} \ket{{\bullet \atop \bullet}{\circ \atop \circ}{\bullet \atop \circ}\cdots {\bullet \atop \circ}}+J_{e,1}\ket{{\circ \atop \circ}{\bullet \atop \bullet}{\bullet \atop \circ}\cdots {\bullet \atop \circ}}\\
    &+J_{e,2}\ket{{\bullet \atop \circ}{\bullet \atop \bullet}{\circ \atop \circ}{\bullet \atop \circ}\cdots {\bullet \atop \circ}}+J_{e_2}\ket{{\bullet \atop \circ}{\circ \atop \circ}{\bullet \atop \bullet}{\bullet \atop \circ}\cdots {\bullet \atop \circ}}\\
    &+J_{e,M-1}\ket{{\bullet \atop \circ}\cdots {\bullet \atop \circ}{\bullet \atop \bullet}{\circ \atop \circ}}+J_{e,M-1}\ket{{\bullet \atop \circ}\cdots {\bullet \atop \circ}{\circ \atop \circ}{\bullet \atop \bullet}} \Bigg),
\end{aligned}
\end{equation}
with the normalization factor $\mathcal{Z}=\sqrt{2\sum_{k=1}^{M-1}J_{e,k}^2}$.
As the $\ket{\phi^\prime_J}$ state only has overlap with states containing one doublon and one hole, it has zero overlap with the scarred states of the first family that have neither of those. As such, any non-trivial dynamics after a quench from this state must come from scarred states of the second family. We can compute this overlap exactly: 
\begin{equation}\label{eq:olap_J}
    |\langle \phi^\prime_J | E_n^\prime \rangle|^2=\frac{\binom{M-2}{n-1}}{2^{M-2}}\frac{\sum_{k=1}^{M-1}J_{e,k}^2}{\left(\sum_{k=1}^{M-1}J_{e,k}^2+2\sum_{k=1}^{M}\omg_k^2 \right)},
\end{equation}
which leads to 
\begin{equation}
    \sum_{n=1}^{M-1}|\langle \phi^\prime_J | E_n^\prime \rangle|^2=\frac{\sum_{k=1}^{M-1}J_{e,k}^2}{\left(\sum_{k=1}^{M-1}J_{e,k}^2+2\sum_{k=1}^{M}\omg_k^2 \right)}.
\end{equation}
It is important to notice here that it is not the $\omega_k$ that enter this equation but their counterparts $\omg_k$, with the mean removed. So if we draw the $J_{e,k}$ and $\omega_k$ from the same distribution $[\Delta-\delta,\Delta+\delta]$, if $\Delta \gg \delta$ then we will have that $J_{e,k}^2 \gg \omg_k^2$ as the former is at the scale of $\Delta^2$ but the latter at the scale of $\delta^2$.
In that case, the state $\ket{\phi^\prime_J}$ will have total overlap of order $\left(1+\frac{2M\delta^2}{(M-1)\Delta^2}\right)^{-1}$ on the QMBSs of the second family, making these states the only relevant ones.

Similarly, we can also define 
\begin{equation}
\begin{aligned}
    \ket{\phi_J}=&\frac{1}{\mathcal{Z}}\sum_{k=1}^M J_{e,k}(\ket{\mathbf{HD}}_{\Lambda_k}\hspace{-0.09cm}{+}\ket{\mathbf{DH}}_{\Lambda_k})\hspace{-0.1cm}\otimes\hspace{-0.1cm}\ket{{\circ \atop \bullet}\cdots {\circ \atop \bullet}}_{\overline{\Lambda}_k} \\
    =&\frac{1}{\mathcal{Z}}\Bigg( J_{e,1} \ket{{\bullet \atop \bullet}{\circ \atop \circ}{\circ \atop \bullet}\cdots {\circ \atop \bullet}}+J_{e,1}\ket{{\circ \atop \circ}{\bullet \atop \bullet}{\circ \atop \bullet}\cdots {\circ \atop \bullet}}\\
    &+J_{e,2}\ket{{\circ \atop \bullet}{\bullet \atop \bullet}{\circ \atop \circ}{\circ \atop \bullet}\cdots {\circ \atop \bullet}}+J_{e_2}\ket{{\circ \atop \bullet}{\circ \atop \circ}{\bullet \atop \bullet}{\circ \atop \bullet}\cdots {\circ \atop \bullet}}\\
    &+J_{e,M-1}\ket{{\circ \atop \bullet}\cdots {\circ \atop \bullet}{\bullet \atop \bullet}{\circ \atop \circ}}+J_{e,M-1}\ket{{\circ \atop \bullet}\cdots {\circ \atop \bullet}{\circ \atop \circ}{\bullet \atop \bullet}} \Bigg).
\end{aligned}
\end{equation}
This state obeys the same $|\langle \phi_J | E_n^\prime \rangle|^2$ as Eq.~(\ref{eq:olap_J}), but with
\begin{equation}
\langle \phi_J | E_n^\prime \rangle=(-1)^{M-1-n}\langle \phi^\prime_J | E_n^\prime \rangle.
\end{equation}
If the $J_e$ only have a small amount of disorder, we can also look at the homogeneous state 
\begin{equation}
\begin{aligned}
    \ket{\phi}=\frac{1}{\mathcal{N}}\Bigg(&\ket{{\bullet \atop \bullet}{\circ \atop \circ}{\circ \atop \bullet}\cdots {\circ \atop \bullet}}+\ket{{\circ \atop \circ}{\bullet \atop \bullet}{\circ \atop \bullet}\cdots {\circ \atop \bullet}}\\
    +&\ket{{\circ \atop \bullet}{\bullet \atop \bullet}{\circ \atop \circ}{\circ \atop \bullet}\cdots {\circ \atop \bullet}}+\ket{{\circ \atop \bullet}{\circ \atop \circ}{\bullet \atop \bullet}{\circ \atop \bullet}\cdots {\circ \atop \bullet}}\\
    +&\ket{{\circ \atop \bullet}\cdots {\circ \atop \bullet}{\bullet \atop \bullet}{\circ \atop \circ}}+\ket{{\circ \atop \bullet}\cdots {\circ \atop \bullet}{\circ \atop \circ}{\bullet \atop \bullet}} \Bigg).
\end{aligned}
\end{equation}
Due to its homogeneous weights, it requires less fine-tuning to prepare. It is well suited for the situation considered before, where  $[\Delta-\delta,\Delta+\delta]$ with $\Delta \gg \delta$.

Finally, for experimental implementations, it is preferable to use an initial state that requires fewest gates to prepare. Because of this, in the main text we have considered the simpler cousin $\ket{\phi_L}$ of the state $\ket{\phi}$:
\begin{equation}
        \ket{\phi_L}=\frac{1}{\mathcal{N}}\Bigg(\ket{{\bullet \atop \bullet}{\circ \atop \circ}{\circ \atop \bullet}\cdots {\circ \atop \bullet}}+\ket{{\circ \atop \circ}{\bullet \atop \bullet}{\circ \atop \bullet}\cdots {\circ \atop \bullet}}\Bigg).
\end{equation}
For this state, the overlap with the scarred states of the second family is 
\begin{equation}\label{eq:olap_L}
    |\langle \phi_L | E_n^\prime \rangle|^2=\frac{\binom{M-2}{n-1}}{2^{M-2}}\frac{J_{e,1}^2}{\left(\sum_{k=1}^{M-1}J_{e,k}^2{+}2\sum_{k=1}^{M}\omg_k^2 \right)}.
\end{equation}
It is important to note that $\ket{\phi^\prime_J}$, $\ket{\phi_J}$, $\ket{\phi}$ and $\ket{\phi_L}$ only have overlap with states with exactly one hole and one doublon. As a consequence, they are completely orthogonal to scarred states of the first family which only have singlets and triplets.

\begin{figure}[tb]
\centering
\includegraphics[width=\linewidth]{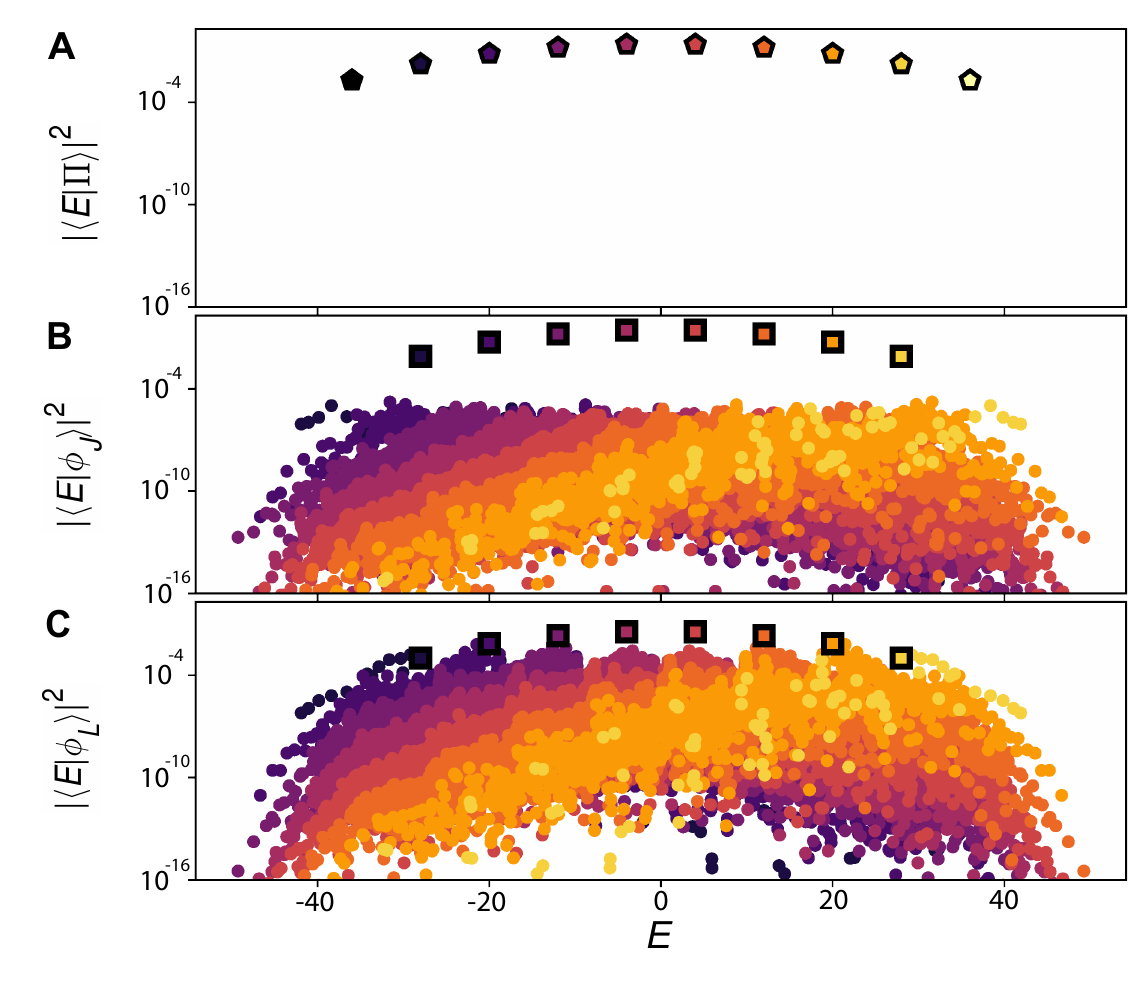}
\caption{{\bf Support of special initial states.} Overlap between the $\ket{\Pi}$, $\ket{\phi_J}$ and $\ket{\phi_L}$ states and the eigenstates of the model in Eq.~(\ref{eq:model}). Scarred states of the first family are denoted by pentagons while those of the second family are denoted by squares. $\ket{\Pi}$ only has overlap on the scarred eigenstates of the first family, while $\ket{\phi_J}$ and $\ket{\phi_L}$ have no overlap on them. For both of these states, the scarred states of the second family dominate. Data is for $N=18$, $J_a=4$ and $J_{e,k}\in [4,4.5]$, $\omega_k\in [0.5,1.5]$ drawn from a uniform distribution.
}
\label{fig:olap}
\end{figure}

\begin{figure}[tb]
\centering
\includegraphics[width=0.95\linewidth]{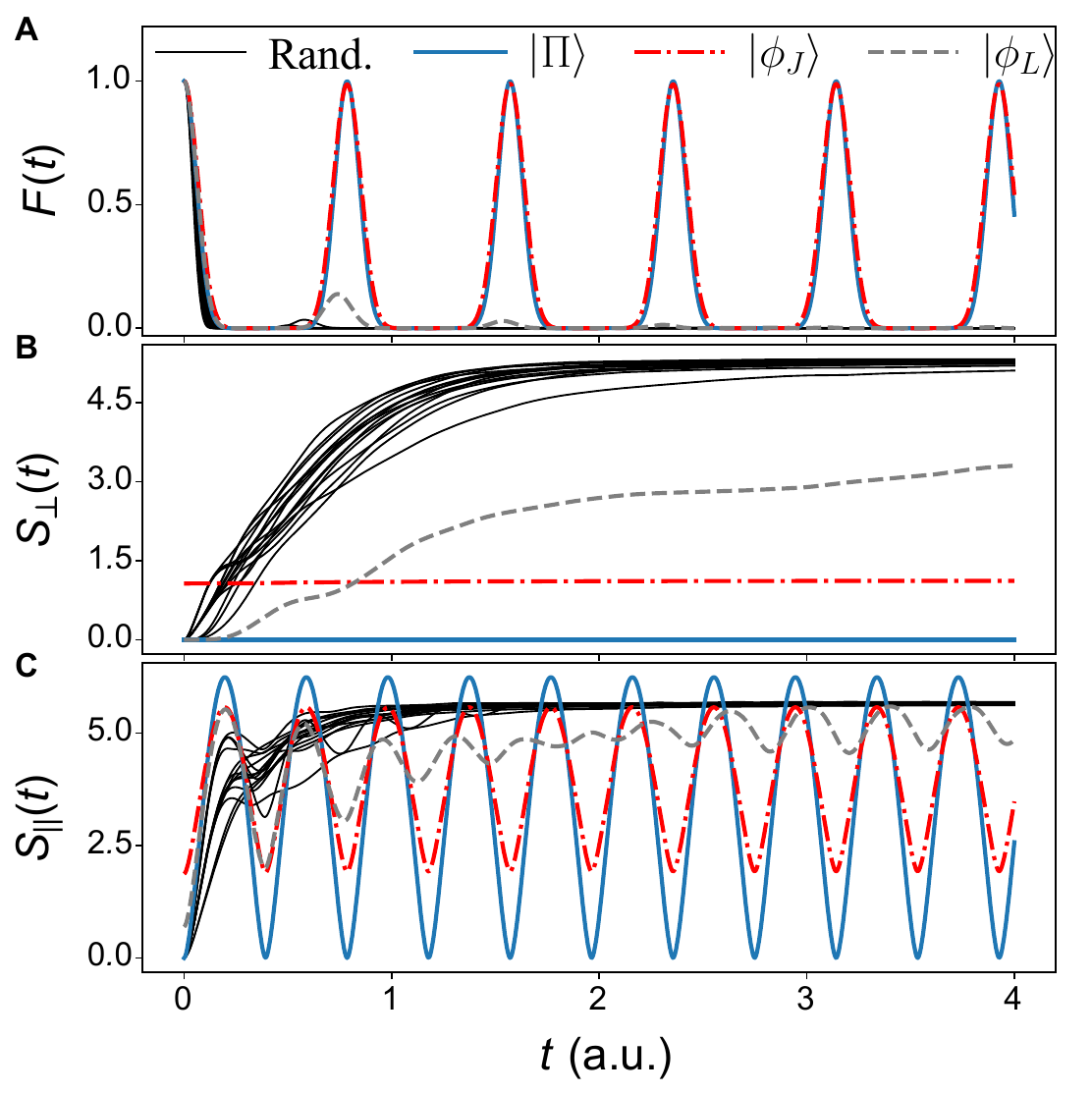}
\caption{{\bf Dynamics from special initial states.} Fidelity and bipartite entanglement entropy over time following the quench from various initial states indicated in the legend. The thin black lines are randomly chosen Fock basis states at half filling. For $\ket{\Pi}$ and $\ket{\phi_J}$ there is no visible growth of entanglement entropy, while for $\ket{\phi_L}$ the growth is strongly suppressed. Data is for $N=18$, $J_a=4$ and $J_{e,k}\in [4,4.5]$, $\omega_k\in [0.5,1.5]$ drawn from a uniform distribution.}
\label{fig:dyn}
\end{figure}

\subsection{Tunability of revivals}
To confirm our analysis above, we have numerically computed the overlap of states $\ket{\Pi}$, $\ket{\phi_J}$ and $\ket{\phi_L}$ with eigenstates of the model in Eq.~(\ref{eq:model}) in Fig.~\ref{fig:olap}. All the states have predominant support on scarred eigenstates, either of the first family ($\ket{\Pi}$ state) or the second family ($\ket{\phi_J}$ and $\ket{\phi_L}$ states). We emphasize that, as scarred states of the first family have no doublons or holons, they are $\emph{exactly}$ orthogonal to $\ket{\phi_J}$ and $\ket{\phi_L}$. 
Thus, persistent revivals from the latter states provide unambiguous evidence for the second type of QMBS. Indeed, Fig.~\ref{fig:dyn}{\bf A} shows that the initial states $\ket{\Pi}$, $\ket{\phi_J}$ and $\ket{\phi_L}$ lead to revivals of the wave function and slow growth of entanglement entropy when compared to random Fock basis states -- clear signatures of scarring. The rainbow nature of $\ket{E_n}$ and $\ket{E^\prime_n}$ is also apparent in the dynamics, as the growth of entropy shows a stark difference between two different cuts, shown in Figs.~\ref{fig:dyn}{\bf B}-{\bf C}.

\begin{figure}[bt]
\centering
\includegraphics[width=\linewidth]{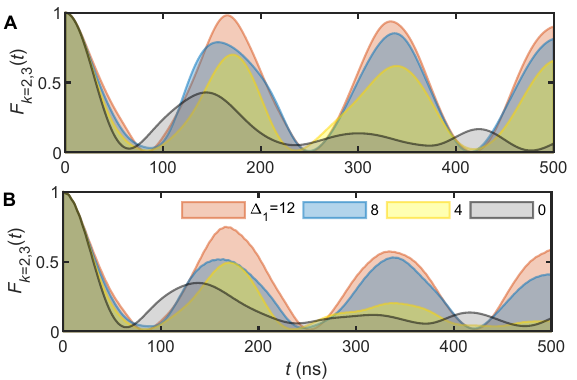}
\caption{
{\bf Tuning of revival by disorder.}
Improving the revivals by modulating the hopping strength on the first site according to $J_{e,1}=J_{e,1}^0+\Delta_1$ for the ideal model ({\bf A}) and for the model with experimental imperfections ({\bf B}) in numerics.
Monotonic increase of revival peaks can be seen as $\Delta_1$ is increased. Data is for $N = 8$
qubits, $J_a = 3.0$ MHz, $J_{e,k} \in[2.0, 3.0]$ MHz, drawn from a uniform distribution. 
} \label{fig:tune2}
\end{figure}

 While the revival fidelity from the $\ket{\phi_L}$ initial state in Fig.~\ref{fig:dyn} is not particularly high, we can leverage the tunability of the second family of scarred states to enhance it. Indeed, from Eq.~(\ref{eq:olap_L}) it directly follows that the projection of $\ket{\phi_L}$  state on the set of scarred eigenstates  $\ket{E^\prime_n}$ is given by
\begin{equation}\label{eq:olap_L_tot_SM}
    \sum_{n=1}^{M-1}|\langle \phi_L | E_n^\prime \rangle|^2=\frac{J_{e,1}^2}{ \sum_{k=1}^{M-1}J_{e,k}^2{+}2\sum_{k=1}^{M}\omg_k^2  }.
\end{equation}
Thus, in the limit of $J_{e,1}{\to}\infty$, we recover perfect revivals. 
To illustrate the effect of $J_{e,1}$ on the dynamics, in Fig.~\ref{fig:tune2} we compute the fidelity dynamics as $J_{e,1}$ is modulated by an amount $\Delta_1 \in [0,12]$, both in the ideal model and the model with experimental imperfections. As expected, for the ideal model the revivals are close to perfect for $\Delta_1=12$. Nevertheless, even after including the imperfections present in the device and realistic values of parameters, we still see a substantial improvement in the first revival peak. We emphasize that the model remains chaotic for the chosen range of $\Delta_1$ values.

\section{Entanglement entropy of scarred eigenstates}\label{app:entanglement}

Due to the su(2) algebra, the first family of scarred eigenstates have the structure of angular momentum eigenstates. As a result, they have entanglement entropy scaling as $\propto \log M$. In fact, it is straightforward to compute their entanglement entropy analytically. Let us assume that $M$ is even and equal to $2R$. For simplicity, we will concentrate on the state with $n=R$ which has exactly zero energy, as it has the highest entanglement entropy among all scarred states. It is easy to decompose the state $\ket{E_R}$ as
\begin{equation}
\begin{aligned}
   \ket{E_R}=& \binom{2R}{R}^{-1/2}\ket{R,R} \\
   =&\binom{2R}{R}^{-1/2}\sum_{k=0}^R \ket{R-k,k}\otimes \ket{k,R-k}\\
  =&\sum_{k=0}^R \frac{\binom{R}{k}}{\sqrt{\binom{2R}{R}}} \ket{\psi_{1,k}}\otimes \ket{\psi_{2,k}},
\end{aligned}
\end{equation}
where $\ket{\psi_{1,k}}$, $\ket{\psi_{2,k}}$ are the normalized versions of $\ket{R-k,k}$ and $\ket{k,R-k}$, respectively. From the last expression, we recognize the prefactors as the Schmidt coefficients. Therefore, the entanglement spectrum has $R+1=M/2+1$ non-zero values with 
\begin{eqnarray}
    p_k=\frac{\binom{R}{k}^2}{\binom{2R}{R}},
\end{eqnarray}
for $k=0,1,\dots,R$. In the large-$M$ limit, one can perform a saddle-point approximation to arrive at the result $S_{1,\perp}=0.5+0.5\log(\pi M/8)$, demonstrating the logarithmic scaling with system size.

\begin{figure}
\centering
\includegraphics[width=\linewidth]{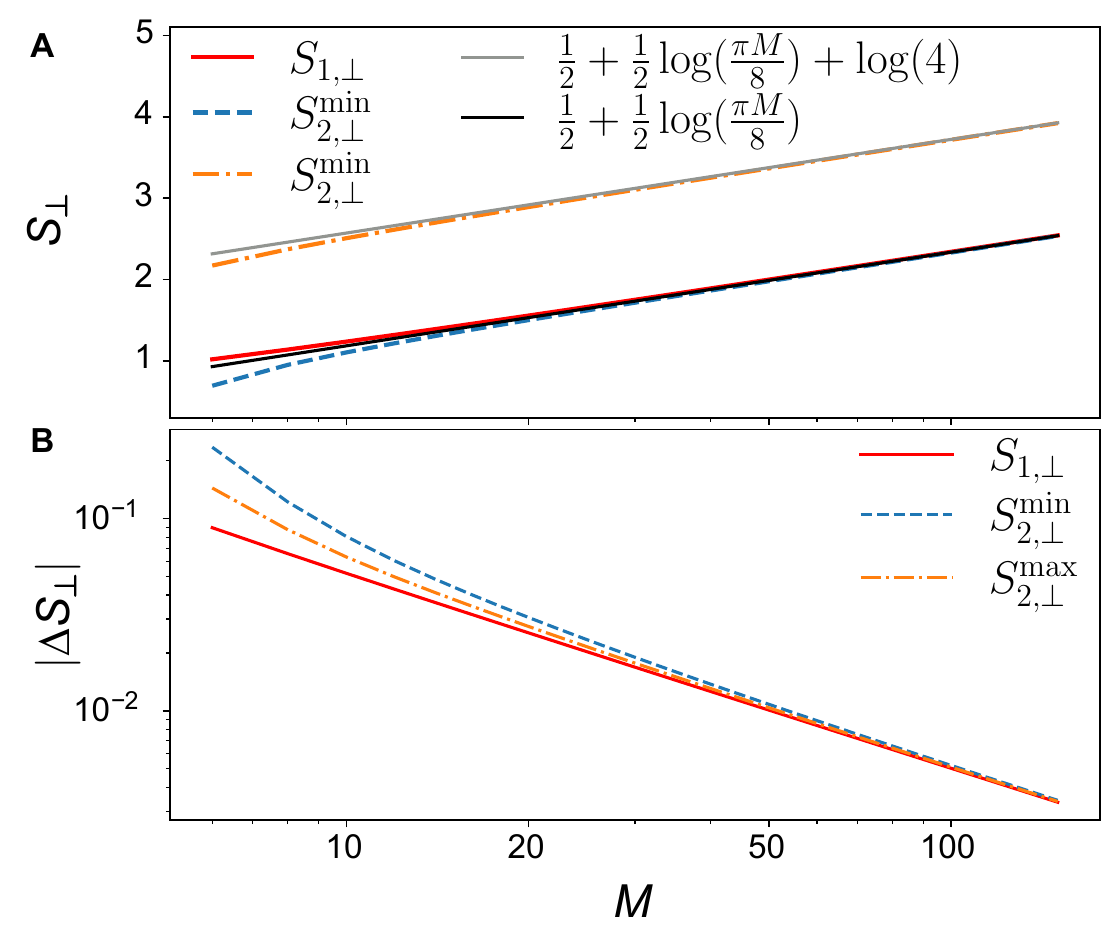}
\caption{{\bf Asymptotic entanglement entropy.} ({\bf A}) Entanglement entropy computed from the analytical formula of the entanglement spectrum. The gray and black lines represent the expected large-$M$ limit and they are in good agreement with the data already for $N\approx 20$. ({\bf B}) Difference between the data and the large-$M$ prediction. For all three cases, the difference approximately decays as $\propto 1/M$.}
\label{fig:S_analytical}
\end{figure}

For the second family of scars, the computation is more arduous as the entanglement entropy depends on the disorder realization. Here we provide an exact computation for a few extremal cases. While we do not have proof that these are the cases with maximum and minimum entanglement entropy, they match the results of our numerical optimizations.
First, we can focus on the $|J_{e,k}|\gg|\omg_k|$ case, in which we assume that the  $\omg_k$ are negligible. Let us also set all $J_{e,k}=1$ equal to 1, except for the middle one which we set to $J_{e,R}=J_e$.
We can then find the Schmidt decomposition of this state as
\begin{equation}
\begin{aligned}
  \ket{E^\prime_R}=&\frac{J_e}{2\mathcal{N}_R} \sum_{k=0}^{R-1} \ket{R{-}1{-}k,k}\ket{\mathbf{D}}\otimes \ket{\mathbf{H}}\ket{k,R{-}1{-}k} \\
   +&\frac{J_e}{2\mathcal{N}_R}\sum_{k=0}^{R-1} \ket{R{-}1{-}k,k}\ket{\mathbf{H}}\otimes \ket{\mathbf{D}}\ket{k,R{-}1{-}k}\\
   +&\frac{1}{2\mathcal{N}_R}\sum_{k=0}^{R-2}\ket{\tilde{\psi}_{1,k}} \otimes \ket{k{+}1,R{-}1{-}k}\\
   +&\frac{1}{2\mathcal{N}_R}\sum_{k=0}^{R-2}\ket{k{+}1,R{-}1{-}k}\otimes \ket{\tilde{\psi}_{2,k}},
\end{aligned}
\end{equation}
with
\begin{align}
    \ket{\tilde{\psi}_{1,k}}&{=}\sum_{j=1}^{R{-}1}\left(\ket{\mathbf{DH}}_{\Lambda_j}{+}\ket{\mathbf{HD}}_{\Lambda_j}\right)\otimes\ket{R{-}2{-}k,k}_{\overline{\Lambda}_j} \\
    \ket{\tilde{\psi}_{2,k}}&{=}\sum_{j=R+1}^{M-1}\left(\ket{\mathbf{DH}}_{\Lambda_j}{+}\ket{\mathbf{HD}}_{\Lambda_j}\right)\otimes\ket{R{-}2{-}k,k}_{\overline{\Lambda}_j},
\end{align}
where ${\Lambda_j}$ denotes sites $j$ and $j+1$ while ${\overline{\Lambda}_j}$ denote all other sites in the same half-system. To find the Schmidt coefficient the only step left is to normalize each ket in the decomposition. We already see that we have at most $4R-2=2M-2$ non-zero coefficients, showing that a state of this form can have, at most, entanglement growing as $\log M$.

Let us first write down $\mathcal{N}_R$ from Eq.~\eqref{eq:Nn} as 
\begin{equation}
    \mathcal{N}_R=\frac{1}{\sqrt{2}}\sqrt{\binom{2R-2}{R-1}\left(J_e^2+M-2\right)}.
\end{equation}
Consequently, the contribution to the entanglement spectrum in the first two sums is identical and given by 
\begin{equation}
    p^{DH}_k=p^{HD}_k=\frac{J_e^2}{2\left(J_e^2+M-2\right)}\frac{\binom{R-1}{k}^2}{\binom{2R-2}{R-1}}.
\end{equation}
Similarly, the third and fourth sums have the same coefficients given by 
\begin{equation}
p^{1}_k=p^{2}_k=\frac{(R-1)}{\left(J_e^2+M-2\right)}\frac{\binom{R-2}{k}\binom{R}{k+1}}{\binom{2R-2}{R-1}}.
\end{equation}
The maximal entanglement entropy is obtained for $J_e=J_e^\star$ and scales as $\sqrt{M}$. In that case, we find numerically that $S^\mathrm{max}_{2,\perp}=S_{1,\perp}+\log 4$ in the large-$M$ limit. It is easy to understand how this additive factor can appear. For the first family of scarred states, we have $R+1$ non-zero values in the entanglement spectrum, while in this case, we have $4R-2$. For $R$ very large, this is a fourfold increase in the number of non-zero values. As they have a similar distribution, this leads to a simple additive factor of 4 due to the log involved in the calculation. 

The case with minimal entanglement entropy is in the limit of a single $J_{e,k}$ (not the middle) being much larger than all other ones. For simplicity, let us consider $J_{e,k}=\delta_{1,k}$. Then the state can be decomposed as
\begin{equation}
\begin{aligned}
  \ket{E^\prime_R}=\frac{1}{\sqrt{2\binom{2R-2}{R-1}}}\sum_{k=0}^{R-2}&\Big[\left(\ket{\mathbf{DH}}{+}\ket{\mathbf{HD}}\right)\otimes\ket{R{-}2{-}k,k}\Big]\\
  &\otimes \ket{k{+}1,R{-}1{-}k}.
\end{aligned}
\end{equation}
This gives us only $R-1$ Schmidt values and the entanglement spectrum can be written down as 
\begin{equation}
p_k=\frac{\binom{R-2}{k}\binom{R}{k+1}}{\binom{2R-2}{R-1}}.
\end{equation}
In the limit of large $M$, we recover the same result as for the scarred state of the first family $S^\mathrm{min}_{2,\perp}=S_{1,\perp}=0.5+0.5\log(\pi M/8)$. This can, once again, be understood simply from the number of nonzero values in the entanglement spectrum as their distribution is similar in both cases. As they have, respectively, $R-1$ and $R+1$ such values, they become identical at the leading order in the large $M$ limit.

\begin{figure}
\centering
\includegraphics[width=0.85\linewidth]{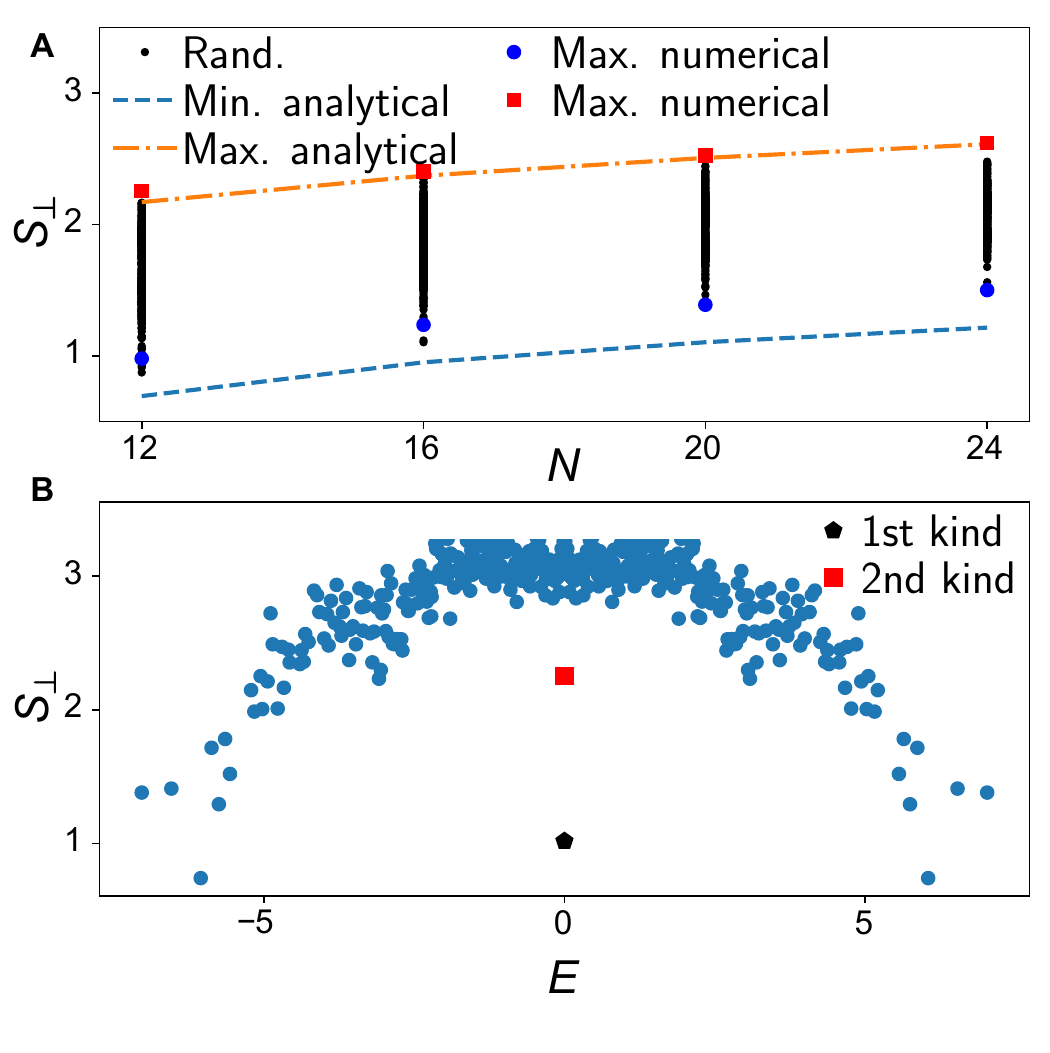}
\caption{{\bf Entanglement entropy for different disorder realizations.} ({\bf A}) Entanglement entropy of the scarred state of the second family with $E=0$ for random realizations, for the analytical formulas, and from numerical minimization/maximization. As $N$ increases, the analytical and numerical results converge. ({\bf B}) Entanglement entropy of eigenstates in the $Q=0$ sector for $N=12$ in the parameter regime was found to maximize the entanglement entropy of the scarred state of the second family. The non-scarred eigenstates concentrate around an arc, as is typical of chaotic systems. This shows that high entanglement entropy can be obtained without getting close to a fine-tuned integrable point where all parameters are identical.  }
\label{fig:S_opti}
\end{figure}

For all three cases, we can compute the entanglement entropy efficiently from the analytical form of the entanglement spectrum for systems with hundreds of sites. Fig.~\ref{fig:S_analytical} displays this along with the expected large-$M$ behavior. We find a very good agreement between them already at $M\approx 20$, with the difference between them decreasing as $1/M$.

While we do not have proof that the cases treated are the true maximum and minimum of $S_{2,\perp}$ we now illustrate numerically that they provide excellent bounds  in the large $N$ limit. Fig.~\ref{fig:S_opti} shows the entanglement entropy obtained for multiple random realizations with different ranges of parameters and for numerical minimization and maximization over all $J_{e,k}$ and $\omega_k$ parameters. For the minimum, we find exact agreement between our analytical and numerical results. For the maximum case, we find that in smaller systems, realizations with high disorder can have higher entanglement entropy than our analytical Ansatz. Nonetheless, the difference between the numerical and analytical maxima converges as $N$ gets larger. This is confirmed by looking at the entanglement spectrum. In all states obtained by numerical maximization of entropy, there are exactly $N=4R$ nonzero values in it. This precludes them from being volume-law states. Asymptotically, this will also be equivalent to the $4R-2$ nonzero values in our analytical Ansatz, up to $1/N$ corrections. For these reasons, we believe that the scarred states of the second family cannot be volume-law entangled and that our analytical Ansatz provides an upper-bound in the large $N$ limit.

The maxima of entanglement entropy obtained numerically are also useful to show that it cannot only be reached by fine-tuned cases where many parameters are identical. As these cases might be integrable or have additional symmetries, they are usually not chaotic. Meanwhile, the numerical maxima do not have degenerate parameters and generically show characteristics of ergodic systems. This can be seen for example in the $N=12$ case, for which the entanglement entropy of eigenstates is plotted in Fig.~\ref{fig:S_opti}.  The concentration of points around an arc is typical of systems obeying the ETH.

\section{General symmetry sectors}
\begin{figure}
\centering
\includegraphics[width=0.85\linewidth]{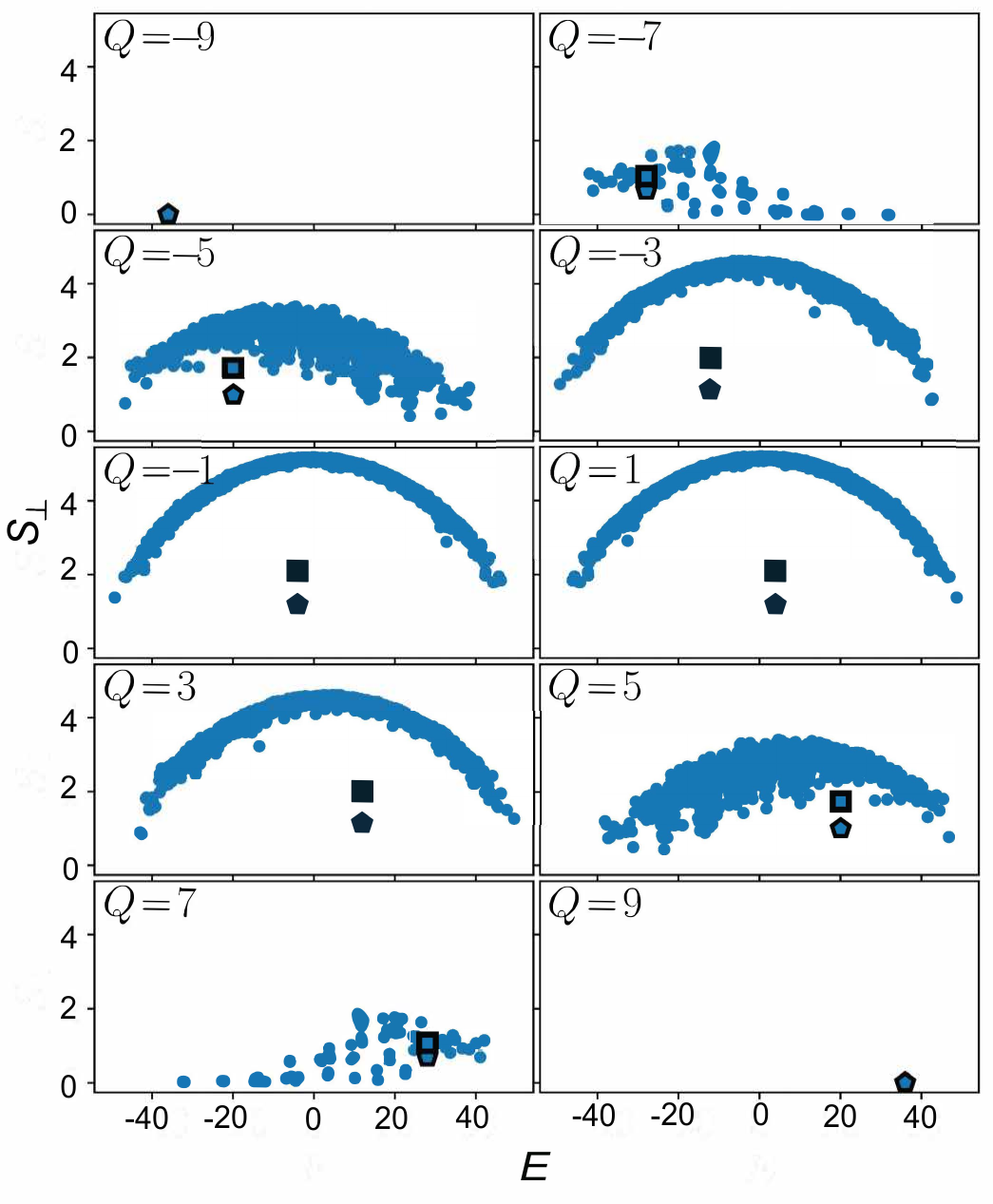}
\caption{{\bf Entanglement entropy in all symmetry sectors.} Entanglement entropy of the eigenstates for a chain with $N=18$, $J_a=4$, $J_{e,k}\in [2,6]$, and $\omega_k\in [1,3]$. Each subplot corresponds to a different sector of $Q$. The scarred states of the first and second families are denoted by pentagons and squares respectively.}
\label{fig:all_Q_hf}
\end{figure}

\begin{figure}
\centering
\includegraphics[width=\linewidth]{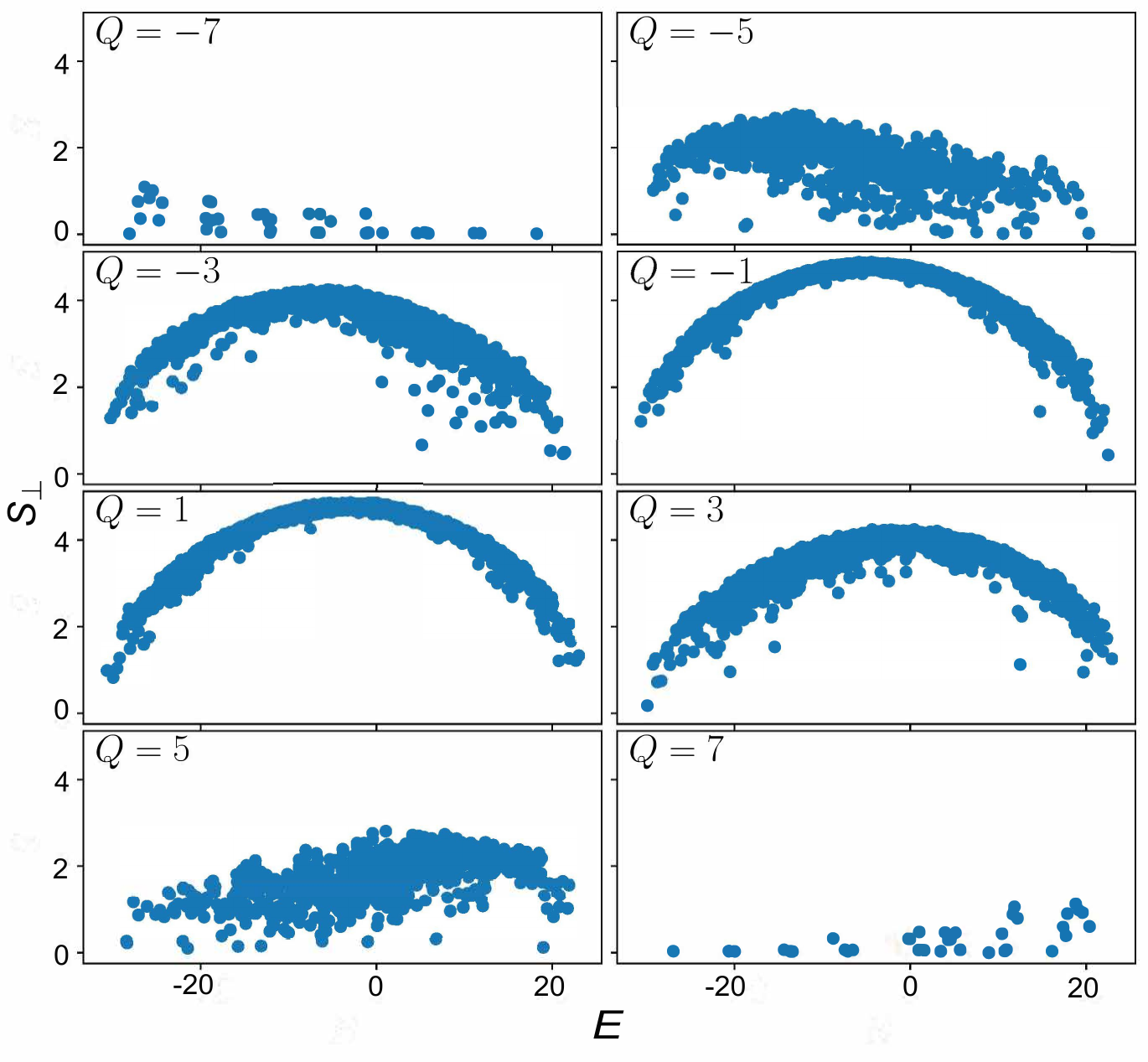}
\caption{{\bf Entanglement entropy away from half-filling.} Entanglement entropy of the eigenstates for a chain with $N=18$ and 7 excitations. Unlike at half-filling (9 excitations), no low-entropy scarred eigenstates can be seen. The system parameters are $J_a=3$, $J_{e,k}\in [2,2.5]$, and $\omega_k\in [0.5,1.5]$.
}
\label{fig:all_Q_fill}
\end{figure}

We have studied different sectors of $\hat{Q}$ symmetry at non-zero magnetization, i.e., at general filling factors in the fermion representation of the model. Fig.~\ref{fig:all_Q_hf} shows the entanglement entropy of eigenstates at half-filling (zero magnetization). This is similar to the data presented in the main text, but we plot individual $\hat Q$ sectors for increased clarity. The scarred states are clearly visible, while the rest of the eigenstates show a narrow arc-like distribution, typical of a chaotic system. By contrast, Fig.~\ref{fig:all_Q_fill} shows a sector away from half-filling, where the same arc is visible. However, in this case, no scarred state is present. Indeed, scarred eigenstates only exist at half-filling, and this can be directly understood from the rainbow scar construction. As the mirror transformation between the two subsystems $\mathcal{M}=\left(\prod_{k=1}^M \hat{\mathbf{d}}_k^x\right)\hat{\mathcal{P}}_{\mathbf{d}\leftrightarrow \mathbf{u}}$ involves a particle-hole exchange, the number of particles in subsystems 1 and 2 must sum to $M$. So by construction, rainbow scars in our model can only exist at half-filling.

\section{Generalization to a non-integrable subsystem Hamiltonian} \label{app:non_integ}

In this section we demonstrate that our results can be generalized to the case in which the subsystem Hamiltonian $\hat{H}_1$ defines a \emph{non-integrable} model. The model presented below is a disordered XY chain with nearest-neighbor and next-nearest-neighbor couplings along the $x$ and $y$ directions. We use the same notation as in the main text and define 
\begin{equation}
    \begin{aligned}
    \hat{H}&=\sum_{j=1}^{M-1}J^1_{x,j}\hat{\mathbf{u}}^x_j \hat{\mathbf{u}}^x_{j+1}+J^1_{y,j}\hat{\mathbf{u}}^y_j \hat{\mathbf{u}}^y_{j+1}\\
    &+\sum_{j=1}^{M-2}J^2_{x,j}\hat{\mathbf{u}}^x_j \hat{\mathbf{u}}^x_{j+2}+J^2_{y,j}\hat{\mathbf{u}}^y_j \hat{\mathbf{u}}^y_{j+2}\\
    &-\sum_{j=1}^{M-1}J^1_{x,j}\hat{\mathbf{d}}^x_j \hat{\mathbf{d}}^x_{j+1}+J^1_{y,j}\hat{\mathbf{d}}^y_j \hat{\mathbf{d}}^y_{j+1}\\
    &-\sum_{j=1}^{M-2}J^2_{x,j}\hat{\mathbf{d}}^x_j \hat{\mathbf{d}}^x_{j+2}+J^2_{y,j}\hat{\mathbf{d}}^y_j \hat{\mathbf{d}}^y_{j+2}\\
    &+\frac{1}{2}\sum_{j=1}^M \hat{\mathbf{u}}^x_j \hat{\mathbf{d}}^x_j+\hat{\mathbf{u}}^y_j \hat{\mathbf{d}}^y_j.
    \end{aligned}
\end{equation}
This Hamiltonian also naturally decomposes into three parts: $\hat{H}_1$ (first two lines) acting on the top row of the ladder, $\hat{H}_2$ (third and fourth line) acting on the bottom row, and $\hat{H}_\mathrm{int}$ linking the two. We can see directly that $\hat{H}_2=-\hat{H}_1$ as the mirror transformation simply swaps the top and bottom row.

\begin{figure}[tb]
\centering
\includegraphics[width=0.8\linewidth]{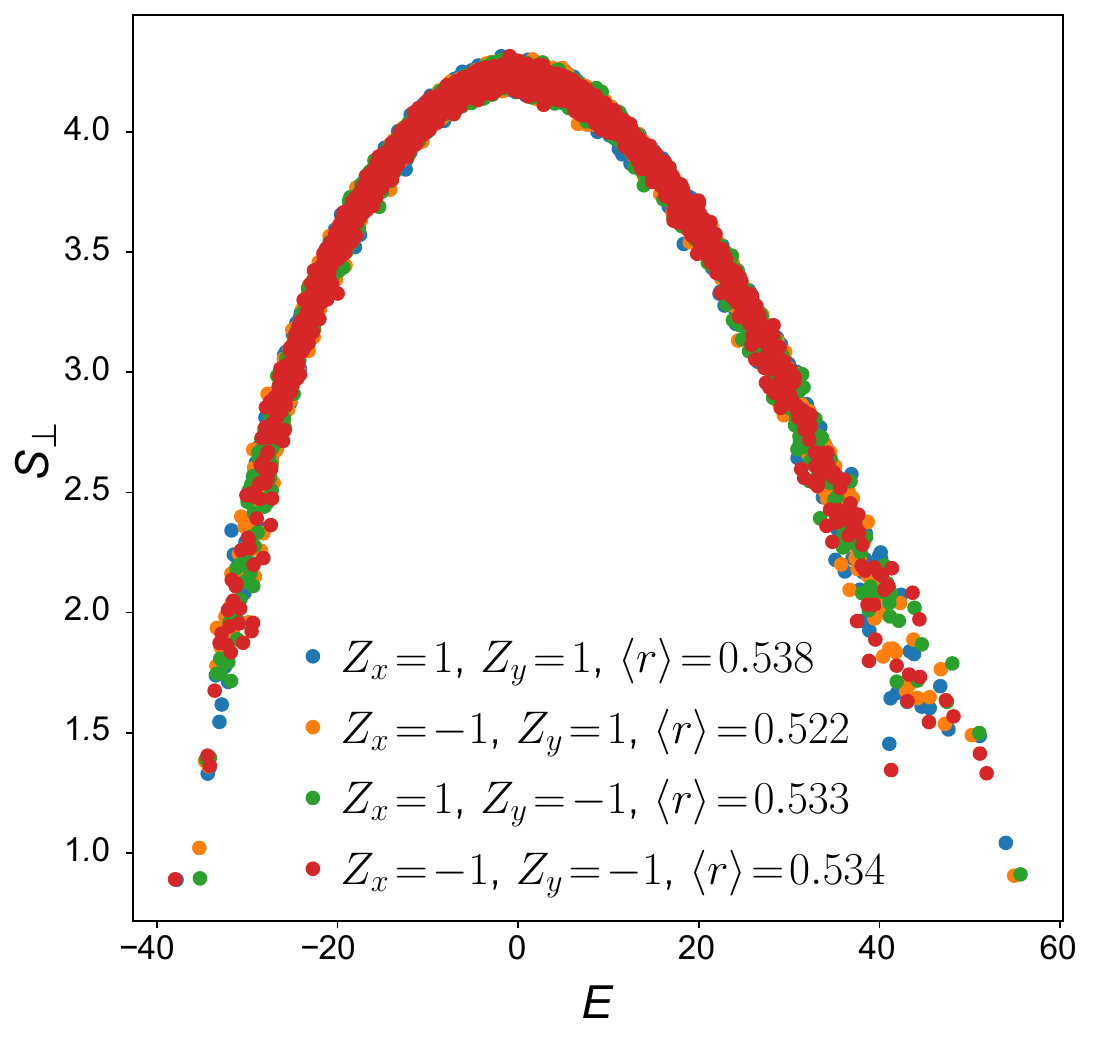}
\caption{{\bf Entanglement entropy in the range-2 XY chain.} Entanglement entropy of the eigenstates with $M=14$, $J^1_{x,k}\in [2.3,2.4]$, $J^1_{y,k}\in [1.2,1.4]$, $J^2_{x,k}=[1.3,1.4]$,
$J^2_{y,k}l\in[2.5,2.7]$. Once the symmetries have been resolved, the level statistics indicate that the system is chaotic (see legend).}
\label{fig:Heis_chaos}
\end{figure}

The Hamiltonian $\hat{H}_1$ has some symmetries linked to $\pi$ pulses along the $x$ or $z$ axes. For an even length $M$, we have two operators $\hat{Z}_x=\prod_{j=1}^M \hat{\mathbf{u}}^x_j$ and $\hat{Z}_z=\prod_{j=1}^M \hat{\mathbf{u}}^z_j$. Once these symmetries have been resolved, the system is clearly chaotic, as is shown in Fig.~\ref{fig:Heis_chaos}.

\begin{figure}
\centering
\includegraphics[width=0.8\linewidth]{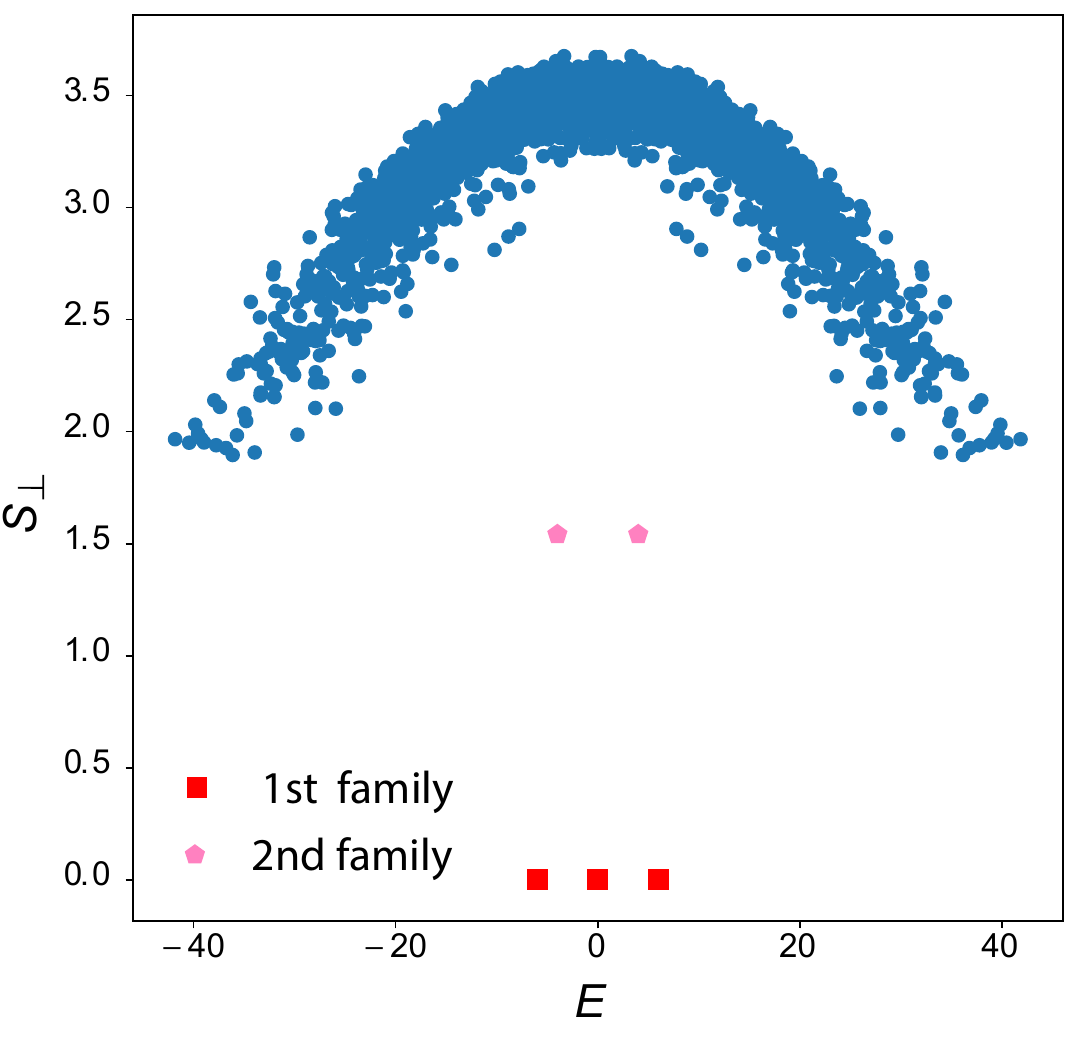}
\caption{{\bf Entanglement entropy in the range-2 XY ladder.} Entanglement entropy of the eigenstates with $N=12$, $J^1_{x,k}\in [2.3,2.4]$, $J^1_{y,k}\in [1.2,1.4]$, $J^2_{x,k}=[1.9,2]$,
$J^2_{y,k}l\in[3.1,3.3]$. The two types of scarred states are well separated in entropy and are clearly distinguishable from the rest of the spectrum.}
\label{fig:Heis_ladder}
\end{figure}

We can now build the rainbow scar state starting from 
\begin{equation}
\ket{\mathbf{I}}=\frac{1}{2^{M/2}}\bigotimes_{k=1}^M \left(\ket{\circ \atop \circ}+\ket{\bullet \atop \bullet}\right).
\end{equation}
By acting with $\hat{Z}_x$ and $\hat{Z}_z$ we can generate two more rainbow scars as 
\begin{align}
\ket{\mathbf{I}_T}&=\frac{1}{2^{M/2}}\bigotimes_{k=1}^M \left(\ket{\circ \atop \bullet}+\ket{\bullet \atop \circ}\right)=\ket{\mathbf{TT}\ldots \mathbf{T}} \\
\ket{\mathbf{I}_S}&=\frac{1}{2^{M/2}}\bigotimes_{k=1}^M \left(\ket{\circ \atop \bullet}-\ket{\bullet \atop \circ}\right)=\ket{\mathbf{SS}\ldots \mathbf{S}}.
\end{align}
These three states have energy respectively $0$, $M$, and $-M$, and have no entanglement entropy for a cut perpendicular to the ladder. We can finally create disordered scarred states (scarred states of the second family) by acting with $\hat{H}_1$ on $\ket{\mathbf{I}_T}$ and $\ket{\mathbf{I}_S}$ to get $\ket{\mathbf{I}^2_T} \propto \hat{H}_1\ket{\mathbf{I}_T}$ and  $\ket{\mathbf{I}^2_S} \propto \hat{H}_1\ket{\mathbf{I}_S}$ which have energy $M-2$ and $2-M$. These states now have doublon-hole pairs $\mathbf{HD}$ and $\mathbf{DH}$ as well as such pairs with a triplet/singlet in between $\mathbf{HSD}$, $\mathbf{HTD}$, $\mathbf{DTH}$ and $\mathbf{DSH}$. The weights of the configuration depend on their location in the ladder and are directly related to the Hamiltonian parameters $J^i_{\alpha,k}$. Similar to the model studied in the main text, these are entangled states with weights that can be tuned. The entropy of the eigenstates is shown in Fig.~\ref{fig:Heis_ladder}
This example is simpler as the symmetries of $\hat{H}_1$ do not appear in $\hat{H}_1$ itself. The Hamiltonian $\hat{H}_1$ is also clearly not integrable, showing that integrability is by no means necessary to build disordered scarred states.

\end{document}